\documentclass[twocolumn]{aastex631}

\received{November 20, 2023}
\revised{January 12, 2024}
\accepted{February 6, 2024}

\submitjournal{AJ}

\shorttitle{Spectroscopic Confirmation of Obscured AGNs from ML}
\shortauthors{Hviding et al.}

\graphicspath{{figures}}

\begin{document}

\title{Spectroscopic Confirmation of Obscured AGN Populations from Unsupervised Machine Learning}

\correspondingauthor{Raphael E. Hviding}
\email{rehviding@arizona.edu}

\author[0000-0002-4684-9005]{Raphael E. Hviding}
\affiliation{Steward Observatory, University of Arizona, 933 North Cherry Avenue, Tucson, AZ 85721, USA}
\affiliation{Max-Planck-Institut für Astronomie, Königstuhl 17, D-69117 Heidelberg, Germany}

\author[0000-0003-4565-8239]{Kevin N. Hainline}
\affiliation{Steward Observatory, University of Arizona, 933 North Cherry Avenue, Tucson, AZ 85721, USA}

\author[0000-0003-4700-663X]{Andy D. Goulding}
\affiliation{Department of Astrophysical Sciences, Princeton University, Princeton, NJ 08544, USA}

\author[0000-0002-5612-3427]{Jenny E. Greene}
\affiliation{Department of Astrophysical Sciences, Princeton University, Princeton, NJ 08544, USA}

\begin{abstract}

We present the result of a spectroscopic campaign targeting Active Galactic Nucleus (AGN) candidates selected using a novel unsupervised machine-learning (ML) algorithm trained on optical and mid-infrared (mid-IR) photometry.
AGN candidates are chosen without incorporating prior AGN selection criteria and are fainter, redder, and more numerous, $\sim$340 AGN\,deg$^{-2}$, than comparable photometric and spectroscopic samples.
In this work we obtain 178 rest-optical spectra from two candidate ML-identified AGN classes with the Hectospec spectrograph on the MMT Observatory.
We find that our first ML-identified group, is dominated by Type I AGNs (85\%) with a $<3\%$ contamination rate from non-AGNs. 
Our second ML-identified group is comprised mostly of Type II AGNs (65\%) with a moderate contamination rate of 15\% primarily from star-forming galaxies.
Our spectroscopic analyses suggest that the classes recover more obscured AGNs, confirming that ML techniques are effective at recovering large populations of AGNs at high levels of extinction.  
We demonstrate the efficacy of pairing existing WISE data with large-area and deep optical/near-infrared photometric surveys to select large populations of AGNs and recover obscured SMBH growth.
This approach is well suited to upcoming photometric surveys, such as Euclid, Rubin, and Roman.

\end{abstract}

% EXplain figures!
% Get Andy to explain ML
% Say specific numbers of spectra
% Confidence in the claim we are making
% Be quantitative about the comparison samples

\keywords{Active galactic nuclei (16), Dimensionality reduction (1943), Galaxy spectroscopy (2171)}

\section{Introduction}\label{sec:intro}

Active galactic nuclei (AGN) are the most luminous sustained phenomena in the universe. 
Powered by accretion onto a central supermassive black hole (SMBH), AGNs are thought to have a significant impact on their host galaxy \citep[for reviews, see][]{fabianObservationalEvidenceActive2012,alexanderWhatDrivesGrowth2012}.
Despite their role in galaxy evolution, a uniform selection methodology for AGNs remains relatively challenging as selected samples of accreting SMBHs using differing techniques return distinct objects, sometimes with little-to-no overlap between criteria \citep[see discussions in][]{hickoxHostGalaxiesClustering2009,padovaniActiveGalacticNuclei2017,hickoxObscuredActiveGalactic2018}.

% Due to the multi-scale and multi-phase nature of SMBH accretion, AGNs emit ubiquitously across the electromagnetic spectrum.
One of the most common forms of AGN identification comes from ultraviolet (UV) and optical photometric selection, which has been used over large areas to search for compact blue emission, a signature of power-law emission from the accretion disk \citep{schmidtQuasarEvolutionDerived1983}.
However, UV and optical photometric selection is sensitive to the obscuration of the central engine. 
Nuclear- or galactic-scale attenuating material can mask the signature of AGN activity, especially for low-luminosity or low-accretion-rate systems. 
Optical spectroscopy, on the other hand, is well suited for determining the {contribution} of an AGN even in the presence of obscuration through the measurement of {enhanced nebular or coronal} narrow emission lines that reveal the ionization signature of SMBH activity {beyond} the extincted nucleus, {i.e}.\ Type II AGNs.
Unobscured AGNs can be detected through the {identification} of broad emission lines, referred to as Type I AGNs.

Large spectroscopic surveys have enabled the detailed characterization of considerable numbers of AGN.
In particular, the Sloan Digital Sky Survey \citep[SDSS;][]{yorkSloanDigitalSky2000} has facilitated the study of SMBH accretion, especially in Type I systems, through the analysis of over 750,000 AGNs {\citep{lykeSloanDigitalSky2020}} detected as a part of the Baryon Oscillation Spectroscopic Survey \citep[BOSS;][]{dawsonBaryonOscillationSpectroscopic2013} and the extended-BOSS \citep[eBOSS;][]{zhaoExtendedBaryonOscillation2016}.
However, these spectroscopic campaigns that target photometrically selected AGNs are biased against low-luminosity and/or obscured systems {which may be below the brightness limit of the instrumentation and/or lack the characteristic compact blue emission that the surveys target}. 
Mass- or magnitude-complete spectroscopic surveys, such as the SDSS Legacy Survey \citep{straussSpectroscopicTargetSelection2002}, are limited to bright populations and to low redshifts $(z\leq0.3)$.

Mid-IR photometric selection of AGNs, conversely, targets re-processed emission from circumnuclear dust heated by SMBH accretion and has traditionally been used as an effective technique for assembling large samples of AGNs at higher levels of obscuration \citep[e.g.][]{lacyObscuredUnobscuredActive2004,sternMidInfraredSelectionActive2005,hickoxLargePopulationMidInfraredselected2007}.
The recovery of obscured systems is of particular importance for exploring their role in galaxy evolution as obscured AGNs may represent a distinct phase of galaxy-SMBH coevolution.
In these systems, material driven into the nucleus contributes both to rapidly feeding the central engine and extincting the optical, UV, and even X-ray signatures \citep{hickoxObscuredActiveGalactic2018}.

The Wide-field Infrared Survey Explorer \citep[WISE;][]{wrightWidefieldInfraredSurvey2010} satellite, which performed an all-sky survey at 3.4, 4.6, 12, and 22 $\mu$m (\textit{W1}, \textit{W2}, \textit{W3}, and \textit{W4} respectively), enabled the selection of tens of millions of AGN candidates \citep[e.g.][]{jarrettSpitzerWISESurveyEcliptic2011,sternMidinfraredSelectionActive2012,mateosUsingBrightUltraHard2012,assefMIDINFRAREDSELECTIONACTIVE2013,assefWISEAGNCatalog2018}.
However, despite its ability to select obscured sources, mid-IR broadband photometry suffers severely from contamination from star-forming galaxies for all but the brightest AGNs \citep{donleyIdentifyingLuminousActive2012,mendezPRIMUSInfraredXRay2013,lamassaSDSSIVEBOSSSpectroscopy2019}.
\citet{hvidingNewInfraredCriterion2022} demonstrated that $>$80\% of spectroscopically-identified AGNs have mid-IR colors that are indistinguishable from those of star-forming galaxies down to a limiting optical brightness. 
In fact, the AGNs that were not selected were dominated by Type II, i.e.\ obscured, and lower-luminosity systems, suggesting there continues to exist a population of AGNs that cannot be selected through mid-IR- or optical-photometric selection alone. 

This bias against obscured sources becomes even more urgent as ongoing and upcoming optical or near-infrared (near-IR) imaging surveys from Hyper Suprime Cam \citep[HSC;][]{miyazakiHyperSuprimeCamSystem2018}, Euclid \citep{laureijsEuclidDefinitionStudy2011}, Rubin \citep{ivezicLSSTScienceDrivers2019}, SPHEREx \citep{doreScienceImpactsSPHEREx2016}, and Roman \citep{spergelWideFieldInfrarRedSurvey2015} will result in large samples of galaxies without accompanying high-resolution ($\mathcal{R}\sim1000$) spectroscopy.
In addition, without any new large-area mid-IR missions with improved spectroscopic or photometric capabilities, WISE-derived datasets remain the best source of mid-IR photometry over large-areas.
It is therefore imperative to explore if existing mid-IR data may be leveraged with the next generation of survey data to develop new AGN selection techniques to recover populations of accreting SMBHs that are systematically missed in traditional selection.

Combining optical imaging and mid-IR photometry hopes to be an effective methodology for selecting obscured and low-luminosity AGNs over large-areas that cannot be selected from a single wavelength regime alone.
In this work we test an unsupervised machine-learning (ML) approach for selecting AGN candidates from deep optical imaging from HSC paired with mid-IR photometry from WISE (Goulding et al., in prep.).
% Critically, the use of optical imaging extends the use of the corresponding mid-IR photometry to lower significance and fainter magnitudes.
Our approach enables us to both make use of the mid-IR data to detect obscured AGNs and the optical color and morphological data to disentangle the AGNs from star-forming galaxy contaminants.
The candidate AGN classes exhibit higher number densities down to fainter magnitudes than comparable methods and it therefore becomes essential to validate our ML approach using deeper optical spectroscopy than previously available.

% We select AGN candidates from our data using an unsupervised machine-learning (ML) algorithm.
% This technique not only efficiently classifies large volumes of data, but trained ML models can immediately be applied to new data.
% In particular, unsupervised ML algorithms do not require labeled data, and can identify patterns and structures without prior knowledge of the labelling of the inputs, potentially being able to identify previously unrecognized groups of objects which may be missed by traditional methods.
% However, it is critical to validate ML predictions to ensure that selected samples are indeed comprised of accreting SMBHs.

We obtain spectroscopic follow-up of ML-identified AGN candidates using the Hectospec spectrograph on the MMT Observatory. 
These observations can confirm the AGN nature of the selected sources, retrieve accurate redshifts, and measure key diagnostic emission-line ratios.
By comparing the properties of the ML-selected AGNs with those of AGNs selected using traditional optical or mid-IR methods, we assess the effectiveness of the unsupervised ML techniques in identifying AGN populations that have been missed by previous surveys.
This spectroscopic follow-up will provide an important test of the ML selection and advance our understanding of the diversity of AGN populations and their role in galaxy evolution.

In Section \ref{sec:ml} we present a brief overview of the ML analysis and the datasets used to inform it that are fully described in Goulding et al.\ (in prep.).
We select our follow-up targets and describe our follow-up spectroscopy in Section \ref{sec:hecto}.
Section \ref{sec:specanalysis} details our spectroscopic analysis, including spectral stacking and emission-line fitting.
Finally, in Sections \ref{sec:disc} and \ref{sec:conc_fut} we present our discussion and conclusions respectively.
Throughout this work we use AB magnitudes for optical photometry, Vega magnitudes for mid-IR photometry, and air wavelengths for optical spectroscopy.
We assume a flat Lambda Cold Dark Matter ($\Lambda$CDM) cosmology with H$_0 = 70$\,km\,s$^{-1}$\,Mpc$^{-1}$ and $\Omega_\textrm{m} = 0.3$.

\section{Sample Selection}\label{sec:ml}

% \begin{itemize}
%     \item Reinforce how our specific combination of data is ideal for finding missing AGN
%     \item Motivate ML and reference Andy's paper
%     \item Summarize the section
% \end{itemize}

Our deep optical imaging from HSC enables us to make greater use of WISE mid-IR photometry.
Typical mid-IR selection criteria require stringent SNR cuts on WISE photomery, such as SNR $>$ $4-5$ in one or more WISE bands \citep{jarrettSpitzerWISESurveyEcliptic2011,sternMidinfraredSelectionActive2012,assefWISEAGNCatalog2018} or even SNR $>$ 10 in \citet{mateosUsingBrightUltraHard2012}.
Through the use of ML techniques to synthesize patterns in our optical and mid-IR datasets, we can incorporate mid-IR data that is at a lower significance than would be used for traditional mid-IR color criteria. 
By leveraging the rich photometric information available from multi-wavelength data, unsupervised ML techniques can help to uncover new and previously missed populations of AGNs over wide areas below the signifiance of existing WISE data.
% Unsupervised ML techniques in particular can provide a powerful alternative for identifying AGN populations in these data-rich environments as they do not rely on prior labeling of the data to be effective.
While the sample selection and ML pipeline is described in full detail in Goulding et al.\ (in prep.), we outline the parent sample in Section \ref{subsec:sample}, summarize the ML procedure to generate our candidate AGN classes in Section \ref{subsec:ml}, and present the properties of our classes in Section \ref{subsec:prop}.

\subsection{Parent Sample}\label{subsec:sample}

% \begin{itemize}
%     \item Describe the HSC and WISE datasets
%     \item In addition reference the SDSS matching
% \end{itemize}

% We briefly summarize the sample selection described in detail in Goulding et al.\ (in prep.).
Goulding et al.\ (in prep.) begin by selecting objects from the Hyper Suprime Cam Subaru Strategic Program \citep[HSC SSP;][]{aiharaHyperSuprimeCamSSP2018}, a large, deep optical imaging survey covering over 1,000 deg$^2$.
The HSC SSP provides imaging in five broadband filters \citep[$g$, $r$, $i$, $z$, and $y$;][]{kawanomotoHyperSuprimeCamFilters2018} with a median five-sigma depth of 25.1 mag and a median seeing of $\sim$0.6 arcseconds.
The data are taken from the HSC SSP Third Data Release \citep{aiharaThirdDataRelease2022}.
Unless otherwise noted, in this work we use the HSC CModel magnitudes, calculated by fitting a linear combination of point-spread-function (PSF) convolved de Vaucouleurs and exponential profiles to the galaxy, to recover the total flux of extended sources.

The superb spatial resolution and limiting surface brightness of the HSC SSP enable morphological characterizations of the target systems.
To generate a non-parametric morphology estimator we also make use of PSF magnitudes that measure the flux of an object assuming it is a point source. 
By taking the difference between the PSF and CModel magnitudes, we measure the deviation of the object from a point source, where a difference larger than zero, i.e.\ where the PSF magnitude is brighter than the CModel magnitude, indicates that the galaxy is more diffuse than a point source\footnote{{A value of less than zero may occur when site conditions at Subaru are better than the fitted PSF shape.}}. 
For a given galaxy, the larger the difference between the PSF and CModel magnitudes, the more flux lies beyond the PSF.

\begin{figure}[ht!]
    \centering
    \includegraphics[width=\columnwidth]{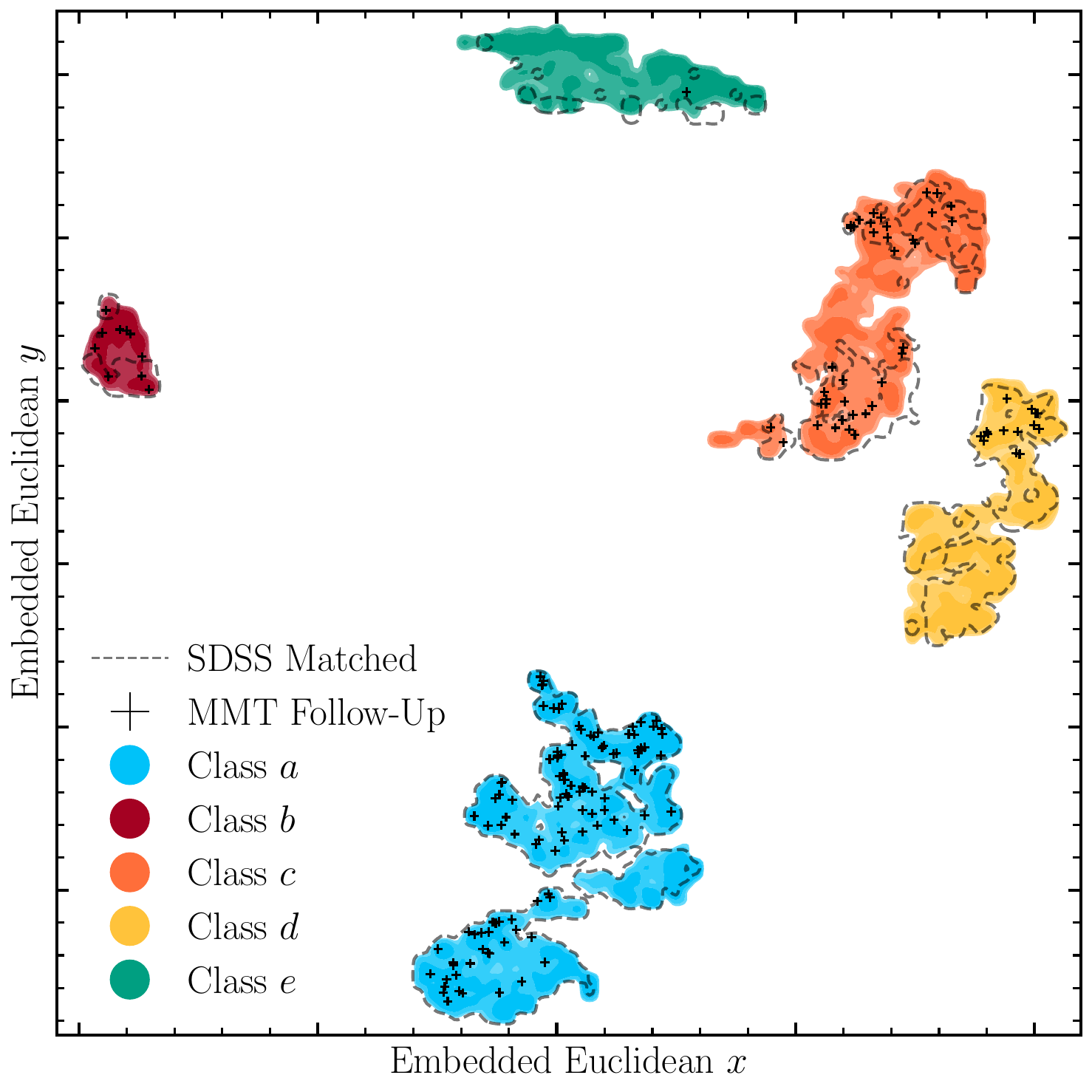}
    \caption{The embedded Euclidean space generated by UMAP from our AGN candidates classes, 3 and 10, after removing stars.
    We color shaded regions encompassing 68\%, 95\%, and 99.7\% of the five subclasses, $a$, $b$, $c$, $d$, and $e$, identified using DBSCAN.
    The dashed contours encompass 99.7\% of the matched SDSS spectroscopy in each subclass.
    Finally, we mark our MMT follow-up targets as plus symbols.
    Throughout this work we refer to subclass $a$ as class I, combine subclasses $b$, $c$, and $d$ into class II, and identify class $e$ as contaminants.\label{fig:classes}}
\end{figure}

\begin{figure*}[ht!]
    \centering
    \includegraphics[width=\textwidth]{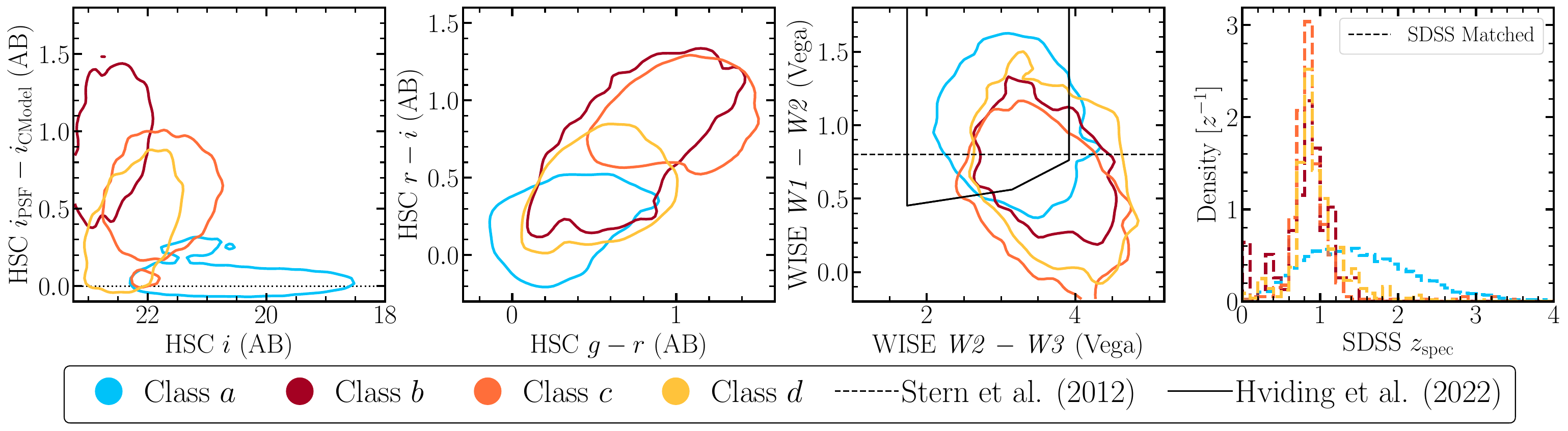}
    \caption{Optical magnitude-morphology (left), optical color-color (center left), mid-IR color-color (center right), and matched SDSS spectroscopic redshift (right) distributions for subclasses $a$ (blue), $b$ (red), $c$ (orange), and $d$ (yellow).
    We plot a contour encompassing 75\% of each class.
    Mid-IR color-color contours are restricted to objects with SNR $>3$ in \textit{W1}, \textit{W2}, and \textit{W3}.
    We plot the \citet[solid]{hvidingNewInfraredCriterion2022} and \citet[dashed]{sternMidinfraredSelectionActive2012} mid-IR selections to highlight regions where typical mid-IR selected AGNs inhabit. {In the leftmost panel we plot a dashed line corresponding to objects with equal PSF magnitudes, corresponding to point sources.} \label{fig:abcd}}
\end{figure*}

For our mid-infrared dataset we make use of both the allWISE \citep{cutriVizieROnlineData2021} and unWISE \citep{schlaflyUnWISECatalogTwo2019} catalogs. While both catalogs leverage data from the WISE cryogenic and NEOWISE \citep{mainzerPreliminaryResultsNEOWISE2011} post-cryogenic surveys, the unWISE catalog provides higher resolution photometry down to fainter fluxes in \textit{W1} and \textit{W2} due the coadding procedure and the preservation of the native WISE resolution. 
We make use of unWISE measurements for \textit{W1} and \textit{W2}, and allWISE measurements for \textit{W3}. 
Due to the low sensitivity in \textit{W4} and the brightness range in our follow-up sample, our analysis does not make use of any \textit{W4} measurements.

Goulding et al.\ (in prep.) matches to all galaxies in SDSS Data Release 16 \citep[DR16;][]{ahumada16thDataRelease2020} spectroscopy using a maximum likelihood approach that takes into account both position and flux.
Through this work we will compare to the properties of the SDSS-matched spectroscopic samples as a comparison to our candidate AGN selections. 
SDSS is representative of optical selection of AGN and therefore highlights the limitations of traditional photometric selection paired with follow-up spectroscopy, which we discuss further in Section \ref{subsec:prop}.

\subsection{Machine Learning}\label{subsec:ml}

Goulding et al.\ (in prep.) begins with the subsample of objects that have detections in all three mid-infrared bands. 
With our eight-band photometry ($g$, $r$, $i$, $z$, $y$, \textit{W1}, \textit{W2}, \textit{W3}), Goulding et al.\ (in prep.) produces the following features:
\begin{itemize}
    \item The 28 possible optical to optical, optical to mid-IR, or mid-IR to mid-IR colors. 
    \item Five non-parametric morphology estimators for each of the optical bands, calculated by taking the difference between PSF and CModel magnitudes. A value larger than zero indicates the galaxy is more diffuse than a point source.
    \item The $r$ band photometry as an absolute flux measurement. 
\end{itemize}
Goulding et al.\ (in prep.) uses a dimensionality reduction algorithm, Uniform Manifold Approximation and Projection\ \citep[UMAP;][]{mcinnesUMAPUniformManifold2018}, along with a clustering algorithm, Density-Based Spatial Clustering of Applications with Noise \citep[DBSCAN;][]{esterDensityBasedAlgorithmDiscovering1996}, to recover 16 classes, two of which are identified as potential AGN classes based on matched SDSS spectroscopy and X-ray datasets and together represent $\sim$13\% of the parent sample.

Following the removal of stellar contaminants using a K-nearest neighbours approach \citep[KNN;][]{fixDiscriminatoryAnalysisNonparametric1951,coverNearestNeighborPattern1967}, Goulding et al.\ (in prep) re-runs UMAP and DBSCAN on the candidate AGN classes to retrieve five total AGN candidate subclasses, $a$, $b$, $c$, $d$, and $e$, which we present in the embedded Euclidean space in Figure \ref{fig:classes}.
By using existing SDSS spectroscopy, Goulding et al.\ (in prep.) concludes that class $e$, which makes up $\sim$16\% of the AGN classes, is dominated by non-AGN sources, and label the subclass as contaminants.
We therefore do not consider subclass $e$ for the remainder of this work.

\begin{figure*}[ht!]
    \centering
    \includegraphics[width=\textwidth]{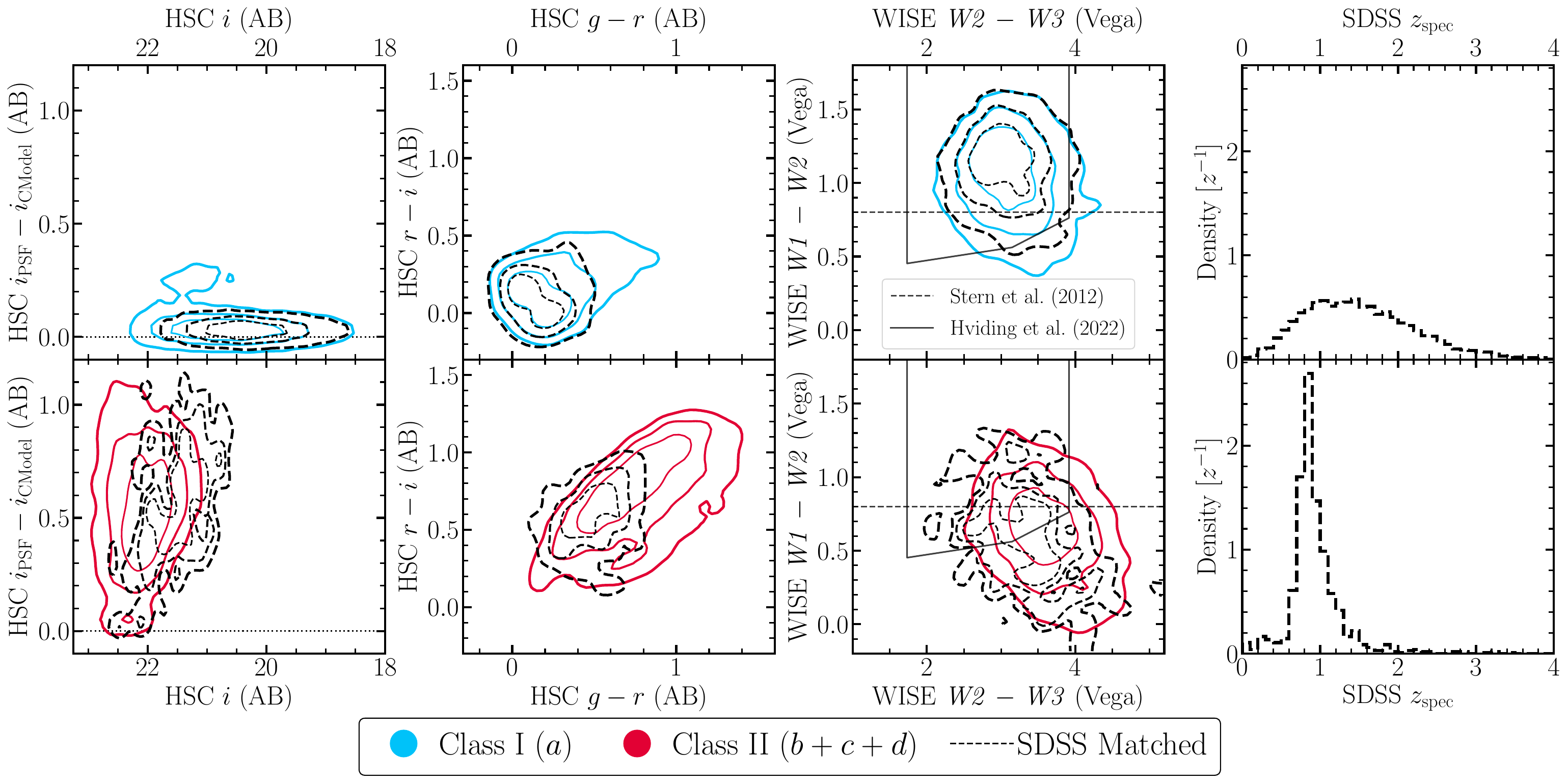}
    \caption{Optical magnitude-morphology (left), optical color-color (center left), mid-IR color-color (center right), and matched SDSS spectroscopic redshift (right) distributions for classes I (top; blue) and II (bottom; red).
    We plot contours encompassing 25\%, 50\%, and 75\% of the classes (solid) and matched SDSS spectroscopy (dashed).
    Mid-IR color-color contours are restricted to objects with SNR $>3$ in \textit{W1}, \textit{W2}, and \textit{W3}.
    We plot the \citet[solid]{hvidingNewInfraredCriterion2022} and \citet[dashed]{sternMidinfraredSelectionActive2012} mid-IR selections to highlight regions where typical mid-IR selected AGNs inhabit.
    Due to SDSS spectroscopic selection, matched objects do not span the entirety of the parameter space of class I and II. In particular SDSS spectroscopy is biased towards the bluest objects in optical color-color space.\label{fig:sdss}}
\end{figure*}

In Figure \ref{fig:abcd} we present the optical magnitude-morphology, optical color-color, mid-IR color-color distributions of subclasses $a$, $b$, $c$, and $d$. 
While our MMT spectroscopy, described in Section \ref{sec:hecto}, is designed to select targets from subclasses $a$, $b$, $c$, and $d$, the majority ($\sim$64\%) of our follow-up targeted subclass $a$ due to its relative brightness.
{In this work we aim to quantify the spectroscopic properties of the ML-selected AGN candidates, including fitting, stacking, and placing constraints on the AGN fractions in our ML-identified classes. 
It is therefore necessary to combine some of the subclasses to have sufficient spectra in each, albeit at the cost of characterizing each subclass individually.}

{We group subclasses $b$, $c$, and $d$ together as these subclasses have less follow-up.
Between subclasses $b$, $c$, and $d$ we observe 63 spectra compared to 114 spectra observed in subclass $a$.
While this risks combining subclasses with different intrinsic properties, we justify the combination due to the similarity in the $i$-band magnitude, WISE color, and SDSS-matched spectroscopic redshifts, as shown in Figure \ref{fig:abcd}.}
For the remainder of this work, we will refer to subclasses $b$, $c$, and $d$ into class II.

\subsection{Class Properties}\label{subsec:prop}

% \begin{itemize}
%     \item Describe Andy's ML
%     \item We want to explain Class I and Class II and motivate them as large sources of AGN
%     \item Then we want to 
% \end{itemize}

Classes I and II represent potential AGN classes identified through photometry alone informed by unsupervised ML. 
If validated, the classes would contain extensive numbers of AGNs over the HSC footprint.
Class I is characterized by a median $i$-band magnitude of $\sim$21 and an on-sky density of $\sim$148\,sources\,$\deg^{-2}$.
On the other hand, class II is, on average, fainter than class I with a median $i$-band magnitude of $\sim$22 but has nearly double the on-sky density with $\sim$235\,sources\,$\deg^{-2}$.
The SDSS Quasar Catalog, for comparison, has an average on-sky density of $\sim$80 AGN\,$\deg^{-2}$ over the SDSS fields with a maximum density of 125 AGN\,$\deg^{-2}$ \citep{parisSloanDigitalSky2018,lykeSloanDigitalSky2020}.
In addition, the 90\% Relability (R90) WISE AGN Catalog, one of the largest and most accurate AGN catalogs in the literature, averages 151 AGN\,$\deg^{-2}$ \citep{assefWISEAGNCatalog2018}.
If their AGN-selection accuracy is high, classes I and II would represent a substantial increase in the number density of AGNs selected via WISE combined with deep optical imaging.

To demonstrate the difference between classes I and II and typical AGN selected samples, we present the optical magnitude-morphology, optical color-color, mid-IR color-color distributions of classes I and II compared to those with existing SDSS spectra along with the corresponding SDSS redshift distributions in Figure \ref{fig:sdss}.
Existing spectroscopic surveys from SDSS and mid-IR selection criteria do not span the full parameter space of our candidate AGN classes.
To highlight the differences in the spectroscopic properties of our ML selected classes and matched SDSS spectroscopy, we compare spectral stacks of these populations in Section \ref{subsubsec:stacksdss}.

\begin{deluxetable*}{c|c|c|l|c|c|rr|rr}
\label{tab:samples}
\tablecaption{Follow-Up Hectospec Fields}
\tablehead{\colhead{Field \#} & \colhead{R.A.} & \colhead{Dec} & \colhead{Obs. Date(s)} & \colhead{Proposal ID} & \colhead{\# Spectra} & \multicolumn{2}{c}{\# Class I} & \multicolumn{2}{c}{\# Class II}}
\startdata
0 & 14:43:25 & $-$00:38:25 & 2020-06-20 & 2020b-UAO-S139 & 193 & 34 & (62.96\%) & 20 (37.04\%) \\
1 & 14:38:15 & $-$00:51:41 & 2020-06-21+22 & 2020b-UAO-S139 & 212 & 36 & (63.16\%) & 21 (36.84\%) \\
2 & 14:39:23 & +00:17:39 & 2020-06-22+23 & 2020b-UAO-S139 & 198 & 38 & (70.37\%) & 15 (27.78\%) \\
3 & 22:47:50 & +00:49:35 & 2020-10-13 & 2020c-UAO-S125 & 127 & 6 & (46.15\%) & 7 (53.85\%)
\enddata
\end{deluxetable*}

In Figure \ref{fig:sdss} we present optical magnitude-morphology, optical color-color, and mid-IR color-color distributions for classes I and II along with the matches for their SDSS matched counterparts demonstrating that SDSS is limited to the brightest, most centrally concentrated, and bluest galaxies for both classes.
For example, $\sim$30\% of class I objects are sufficiently red to have $g-i>0.75$, whereas this is only true for $\sim$15\% of the SDSS-matched subsample. 
For class I this is due to the fact that nearly all SDSS-matched spectroscopy in this class is targeted as part of quasar (90\%) or stellar mapping (10\%) subsurveys within SDSS, BOSS, and eBOSS which target compact blue emission.
% In fact, for class I, 30\% of SDSS matches are from BOSS, 60\% from eBOSS, and 10\% from other SDSS surveys.

For class II the limitations of SDSS-matched spectroscopy is especially noticeable, where $\sim$80\% of {matches are} limited to $i < 22$, while half of class II is fainter than this limit.
While $\sim$50\% of class II galaxies have $g-i>1.5$, only $\sim$15\% of SDSS-matched objects are sufficiently red. 
However, despite having nearly twice the on-sky density as class I, class II has {only 11\%} the number of SDSS matches, driven, in part, by the fact that class I is brighter than class II on average by $\sim$1.43\,mag in the $i$ band ($\sim$3.73 times brighter). 
While matched SDSS spectroscopy for class II are also targeted by quasar subsurveys (20\%), they are primarily observed in eBOSS emission-line galaxy subsurveys (80\%) that target galaxies at $z\simeq0.9$ with strong [\ion{O}{2}] emission
\citep{comparatSDSSIVEBOSSEmissionline2016}.
Therefore class II has over 90\% of matched SDSS spectra from eBOSS, with 10\% from BOSS, and less than a percent from other SDSS surveys. 

%,delubacSDSSIVEBOSSEmission2017,raichoorSDSSIVExtendedBaryon2017}.
% This discrepancy becomes even more pronounced for more diffuse galaxies and highlights the necessity for the deep optical imaging from HSC.

Figure \ref{fig:sdss} emphasizes that existing SDSS spectroscopy has focused on the bluest subset of targets, i.e.\ 
SDSS spectroscopy is therefore not able to inform the AGN-selection accuracy of our candidate classes as it primarily targets bright, blue, and compact galaxies when compared to our selection.
Additional spectroscopic follow-up is therefore necessary to fully span the parameter space of our ML-selected AGN classes.
% Our observations were chosen to span the optical color and morphological distribution for each of the classes.
% We obtain MMT spectroscopy to determine the AGN fraction and type of our ML-selected classes. 

\section{MMT Hectospec Follow Up}\label{sec:hecto}
 
% \begin{itemize}
%     \item Explain the Follow-Up Spectroscopy Sample
%     \item Describe the instrument and data
%     \item Detail the reduction
%     \item Walk through the redshift determination.
% \end{itemize}

To validate the efficacy of the ML pipeline at selecting AGNs, we were awarded two nights in the spring of 2020 to observe five fields\footnote{Due to the outbreak of COVID-19 in early 2020, our observing time was split over two semesters. In addition, due to instrument repairs, we were only able to observe four out of our five fields.} with Hectospec, a multi-object fiber-fed spectrograph installed on the 6.5m MMT Observatory \citep{fabricantHectospecMMT3002005}.
Hectospec is the ideal instrument to follow-up large numbers of photometrically-selected AGN candidates owing to its multiplexing capability to obtain moderate-resolution spectra of up to 300 objects simultaneously over a 1$^\circ$ diameter field using 1.5$''$-diameter fibers.
We utilized the 270\,lines\,mm$^{-1}$ grating, affording us a $\mathcal{R}\sim1000$ over a 3650$-$9200\AA\ wavelength range.
{Targets are selected from across all UMAP identified classes, with a higher priority designated for MMT follow-up targets from classes I and II, which are the focus of this work.}
We select our MMT Hectospec targets in Section \ref{subsec:target}, describe our data reduction in Section \ref{subsec:reduc}, and determine spectroscopic redshifts in Section \ref{subsec:redshift}.

\subsection{Target Selection}\label{subsec:target}

Our target selection was motivated to find fields with relatively high densities of AGN candidates, e.g.\ objects from classes I and II.
In addition, fields were selected to ensure an appropriate density of potential guide stars and standard stars.
Guide stars are selected from Gaia Data Release 2 \citep[DR2;][]{gaiacollaborationGaiaDataRelease2018} to have G-band magnitudes in the range $13 - 15$. Standard stars were selected from SDSS DR16 following the ``Target Selection of F Star Photometric Standards''\footnote{\url{https://www.sdss.org/dr16/algorithms/boss\_std\_ts/}}.
Fiber placements are generated for candidate fields using the  \texttt{xfitfibs}\footnote{\url{https://lweb.cfa.harvard.edu/mmti/hectospec/xfitfibs/}} utility which allows for the ranking of the target galaxies by their class and assigns fibers for the determination of the sky-background.

We observed our four selected fields in queue mode over two semesters in 2020, which are listed in Table \ref{tab:samples}.
Each field was observed for a total of three hours comprised of six 30-minute integrations.
Our targets were drawn from classes I and II while any remaining fibers were assigned to galaxies from  other ML-identified UMAP classes.
We observe a total of 178 galaxies across our ML-selected classes, with 114 ($\sim$64\%) from class I and 63 ($\sim$36\%) from class II: {10, 38, and 15 from subclasses $b$, $c$, and $d$ respectively}.
% In Figure \ref{fig:color} we present the optical magnitude-morphology, optical color-color and mid-IR color-color distributions of our selected targets with respect to our parent HSC sample along with the MMT redshift recovery fraction as a function of optical magnitude which we discuss in Section \ref{subsec:redshift}.
Due to the limit of MMT Hectospec, our spectroscopy is restricted to sources with $i \lesssim 22.5$.
Consequently our analysis of class II will be limited the brightest 75\% of objects.

\subsection{Data Reduction}\label{subsec:reduc}

The Hectospec spectra were reduced using the \texttt{HSRED}\footnote{\url{http://www.mmto.org/hsred-reduction-pipeline/}} reduction pipeline.
The pipeline bias corrects, dark subtracts, and flat fields the data using the provided calibration images.
Each spectrum is extracted, associated with the input catalog objects, and then wavelength calibrated using arc lamp spectra.
{Sky-background and flux corrections are then applied} using the data from the fibers reserved for sky background and F-Star Photometric Standards.
Objects which are observed over multiple nights are coadded after running the \texttt{HSRED} pipeline on each night individually.
{Finally, the spectra are corrected for Milky Way extinction based on the \citet{schlaflyMeasuringReddeningSloan2011} maps assuming a \citet{fitzpatrickCorrectingEffectsInterstellar1999} dust law $(R(V) = 3.1)$}

% Due to fiber errors, we note that two objects, 41227717491846844 and 42296438499078935 were only observed for 1.5 hours on the first and second night of their runs respectively.

\begin{figure}[ht!]
    \centering
    \includegraphics[width=\columnwidth]{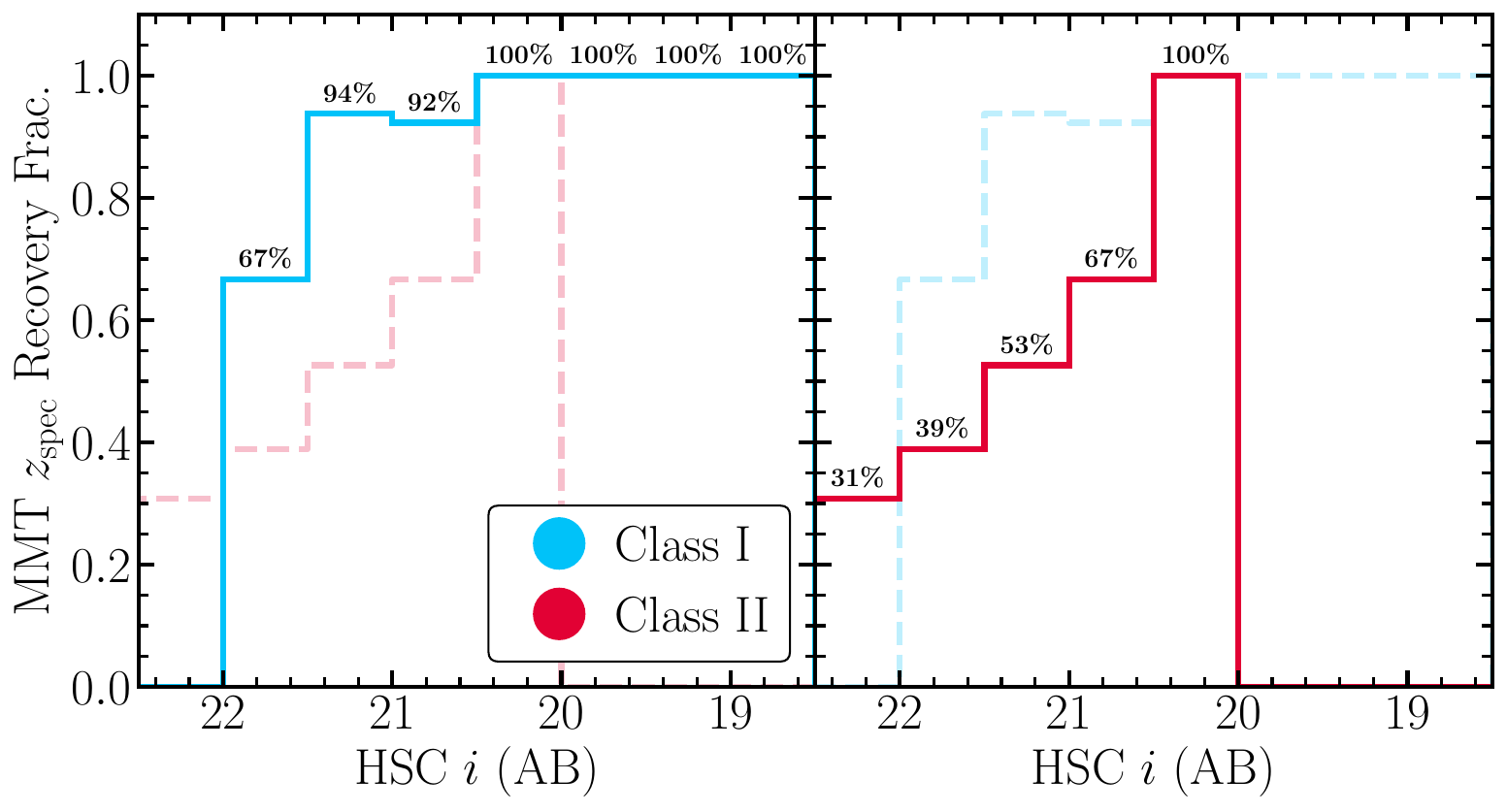}
    \caption{MMT redshift recovery fraction against optical magnitude distributions for classes I (blue; left) and II (red; right).
    We note that our follow-up spectroscopy are limited to $i < 22.5$, which particularly effects class II.\label{fig:redrecover}}
\end{figure}

In addition, 17 of our objects have existing spectroscopy from SDSS, though none of which are in our candidate AGN classes.
As some of our fields did not have sufficient F-stars to construct a flux calibration, we generate and apply an additional calibration to our spectra based on the reobserved targets for the relevant fields.
This is primarily used to retrieve physically motivated continuum fits as described in Section \ref{subsec:emiss}.
We note this has a negligible effect on our subsequent analysis as we are concerned with the detection of broad emission lines or measurement of flux ratios of nearby lines, which are minimally affected by color corrections.

\subsection{Redshift Determination}\label{subsec:redshift}

MMT spectroscopy enables us to accurately determine the redshift of our sample by searching for prominent emission lines, which we describe at the start of Section \ref{sec:specanalysis}, or absorption features such as the Ca H\&K lines or the 4000\AA\ break. 
To determine the redshift of our sample, all reduced spectra are visually inspected to generate an initial estimate for where two or more spectral features can be clearly identified.
We make use of the \texttt{redshifting}\footnote{\url{https://github.com/sdjohnson-astro/redshifting}} code which performs a grid-search using eigenspectrum templates from \citet{boltonSpectralClassificationRedshift2012} to determine redshift from optical spectroscopy \citep{johnsonGalaxyQuasarFueling2018,heltonDiscoveryOriginsGiant2021,johnsonDirectlyTracingCool2022}. 
The spectra are fit over a range of redshifts using the quasar ($0 < z < 3$), galaxy ($0 < z < 1.5$), and star ($-0.005 < z < 0.005$) eigenspectra and we compare the retrieved $\chi^2$ and $\chi^2_\nu$ distributions to refine and improve our visual inspection redshifts.

\begin{figure}[ht!]
    \centering
    \includegraphics[width=\columnwidth]{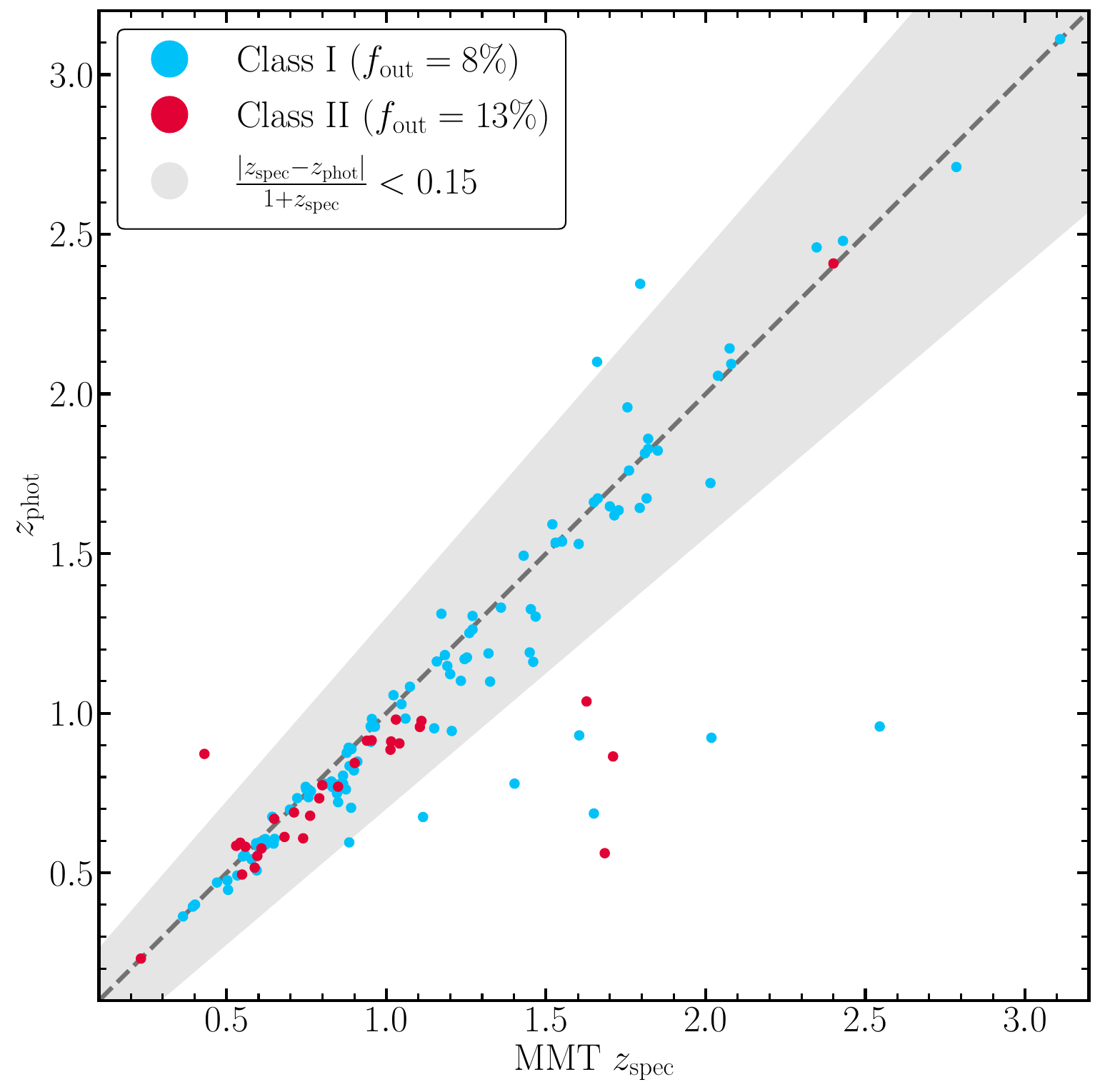}
    \caption{Photometric redshift against spectroscopic redshift for classes I (blue) and II (red).
    \label{fig:zcomp}}
\end{figure}

We retrieve redshifts for 139 ($\sim$78\%) of our class I and II follow-up targets.
In Figure \ref{fig:redrecover} we show the redshift recovery fraction as a function of optical magnitude. 
As expected, we are less successful at identifying redshifts for fainter sources and are most successful at $i < 21.5$, with redshift recovery fractions of 97\% and 60\% for classes I and II respectively.
In addition, we are less able to identify redshifts for class II, especially for the faintest objects, which we attribute to a relative lack of strong emission lines as compared to class I, as discussed in Section \ref{subsec:stack}.
Combined with our spectroscopic limit of $i < 22.5$, our analysis of class II is limited to the brightest 75\% of objects in the class.
In Figure \ref{fig:zcomp} we compare the photometric redshift estimates from Goulding et al.\ (in prep.) designed for AGN and quasar candidates with our recovered spectroscopic redshifts for both classes I and II. Briefly, the combined HSC and WISE photometry for AGN with archival spectroscopy are used to train a set of augmented Random Forest algorithms to construct and assess the accuracy and precision of photometric redshifts for both the unobscured and obscured AGN populations.
They show that these ML$-$based photo$-$zs perform equally well on Type I and Type II AGN alike out to $z \sim 3$ with an average precision of $\delta z / (1+z) \simeq 0.02$ and 0.03, respectively, and that they significantly out-perform the standard HSC \texttt{Mizuki} photometric redshifts that are provided by Subaru$-$HSC \citep{nishizawaPhotometricRedshiftsHyper2020}.
Using our sample of AGN with new redshifts from MMT/Hectospec, we find that both AGN classes have good agreement with moderate outlier fractions $\sim10$\%, reinforcing the efficacy of the photometric redshifts for AGN presented in Goulding et al.\ (in prep.).

\begin{figure}[ht!]
    \centering
    \includegraphics[width=\columnwidth]{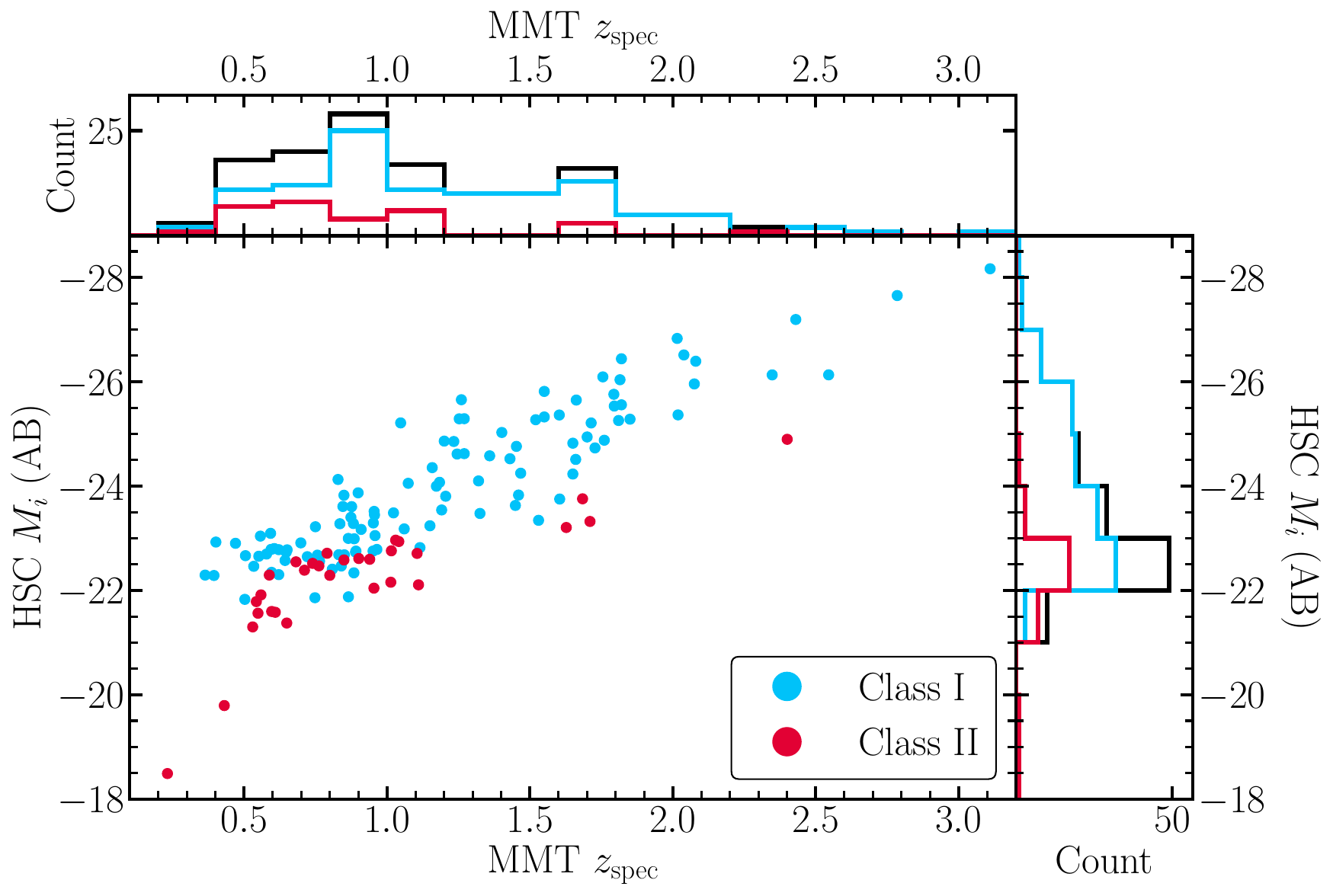}
    \caption{Absolute observed-frame $i$-band magnitude versus spectroscopic redshift for class I (blue) and class II (red) objects with recovered MMT redshifts. The top and right panels present the redshift and magnitude histograms respectively for classes I and II along with an overall histograms in black.\label{fig:mag}}
\end{figure}

In Figure \ref{fig:mag} we plot the absolute magnitude versus spectroscopic redshift for our targets.
Overall class I sources are more luminous and at higher redshifts than those in class II. 
Class I is distributed across redshifts from $z = 0.5-3$.
The redshift distribution for class II is clustered around  $z = 0.5-1$, with potentially a tail out to $z\sim2$, though we are limited by the number of targets with redshift in this class. 
For both classes the distributions are similar to those shown in Figure \ref{fig:sdss} for matched SDSS samples.

\begin{figure*}[ht!]
    \centering
    \includegraphics[width=\textwidth]{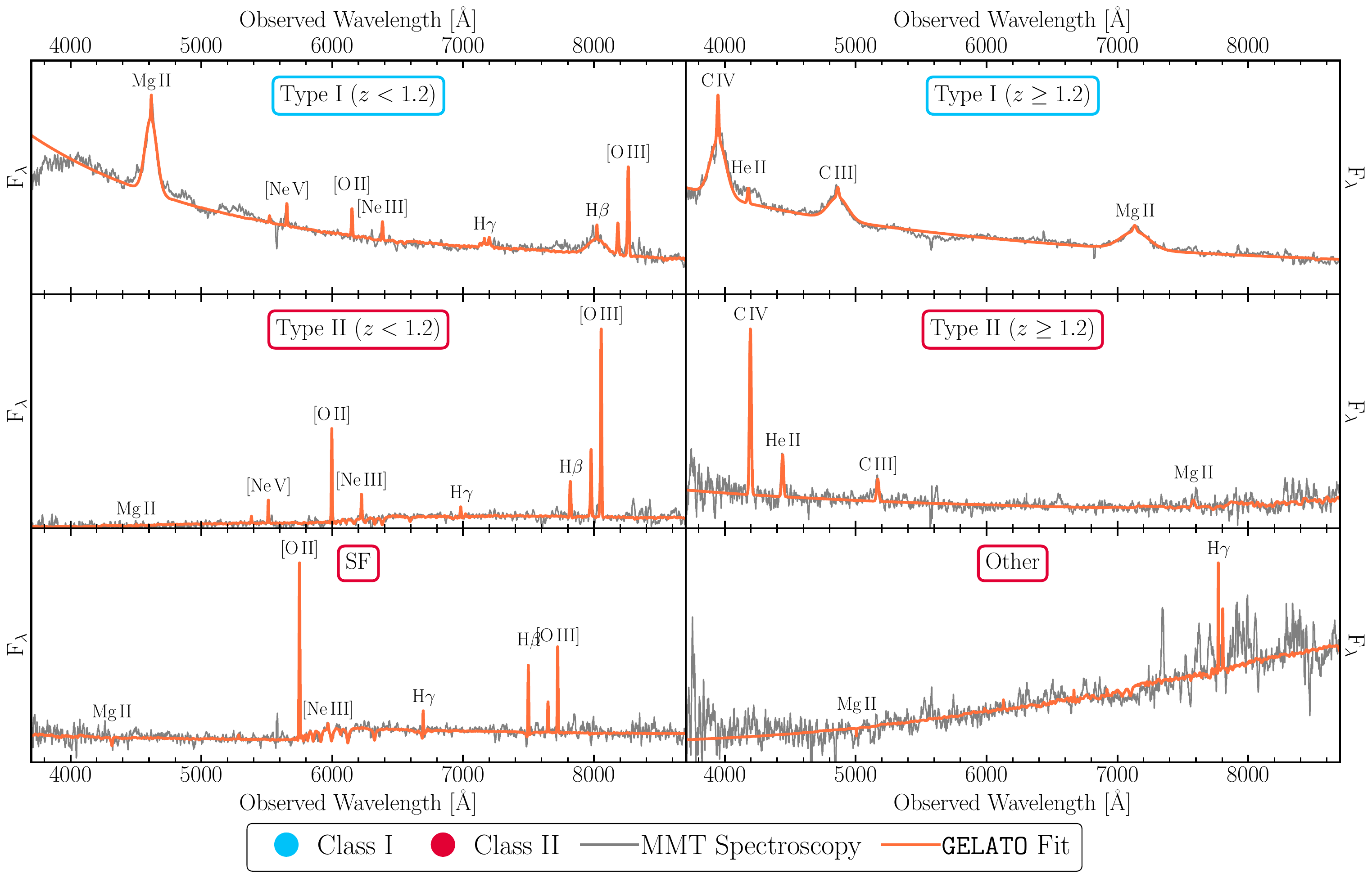}
    \caption{Example MMT spectra from our spectroscopic samples (black) along with the corresponding \texttt{GELATO} fits (orange) for select class I (blue) and class II (red) targets. The objects are chosen across the redshift range of our sample and to be representative of the spectral classifications used in this work. Spectra are smoothed with a boxcar function of width of 10\AA\ for plotting purposes only.
\label{fig:examples}}
\end{figure*}

\section{Spectroscopic Analysis}\label{sec:specanalysis}

% \begin{itemize}
%     \item Detail our spectroscopic fitting analysis
%     \item Start by describing the redshift split in the data
%     \item Highlight GELATO and the show examples
%     \item Detail selection:\
%     \begin{itemize}
%         \item Broad line
%         \item NeV
%         \item TBT Diagram
%         \item UV EL diagram
%     \end{itemize}
%     \item Create probability than an object is an AGN
%     \item Present the stacking analysis
%     \item Get the stacks per magnitude to show overall properties
%     \item Split by optical color and show that extinction can explain the differenceas
% \end{itemize}

In this Section we use our MMT follow-up to determine (a) the AGN fraction per class and (b) the properties of these AGNs in our ML-selected targets.
Out to $z \lesssim 1$ optical spectroscopy enables the characterization of both the [\ion{O}{3}]$\lambda\lambda5008,4960$\AA\ doublet, which is sensitive to the ionization state of nebular gas and to potential AGN-driven outflows, as well as the detection of broad Balmer or \ion{Mg}{2}$\lambda2799$\AA\ line emission.
At intermediate redshifts ($0.5 \lesssim z \lesssim 1.5$) [\ion{Ne}{3}]$\lambda3870$ to [\ion{O}{2}]$\lambda\lambda3730,3727$\AA\ nebular emission-line ratios and the detection of the [\ion{Ne}{5}]$\lambda\lambda3426.85,3346.79$\AA\ coronal emission doublet are diagnostics for AGN activity.
Finally, at the highest redshifts probed by our analysis ($1.5 \lesssim z \lesssim 3$), the strengths and widths of rest-ultraviolet (rest-UV) emission lines such as Ly$\alpha$, \ion{Si}{5}$\lambda1398$\AA, \ion{C}{4}$\lambda1549$\AA, \ion{He}{2}$\lambda1640$\AA, and \ion{C}{3}]$\lambda1909$\AA\ trace SMBH accretion.

To determine the AGN properties and prevalence in our ML-selected candidate AGN classes, in this section we
measure the spectroscopic properties of the 139 galaxies with identified redshifts. 
Emission-line fitting is performed in Section \ref{subsec:emiss} to measure the ionization properties and kinematics in our galaxies to determine the presence of AGN activity and compute the the AGN fraction in our ML selected classes.
We perform spectral stacking to determine average properties for classes I and II (Section \ref{subsec:stack}) and present Balmer decrements to measure galaxy-scale attenuation in Section \ref{subsec:baldec}.

\subsection{Emission-Line Fitting}\label{subsec:emiss}

To fit our spectra we make use of the Galaxy/AGN Emission-Line Analysis TOol (\texttt{GELATO})\citep{GELATOv2.5.2}.
\texttt{GELATO} is designed to retrieve emission-line parameters in optical spectroscopy {by fitting a series of Gaussians for emission lines, Simple Stellar Populations (SSPs) for the underlying continuum, and a power-law continuum\footnote{{We note that \texttt{GELATO} does not include Iron emission templates. While these would be necessary for in-depth study of broad-line quasars in our sample, they are not necessary for the robust detection of broad lines or the characterization of the narrow-line emission in obscured and/or low-luminosity AGN.}}.
Critically \texttt{GELATO} statistically tests for the presence of broad-lines and is therefore ideal for this work.}
We begin by splitting our spectra into two subsamples, a low-redshift ($z < 1.2$) and a high-redshift ($z \geq 1.2$) subsample.
As \texttt{GELATO} uses the Extended MILES stellar library \citep[E-MILES;][]{vazdekisUVextendedEMILESStellar2016} with SSPs that only extend to a wavelength of 1680\AA\, we only fit our high-redshift subsample with a flexible power-law continuum and no SSP templates.

We use \texttt{GELATO} to fit all emission lines listed at the start of Section \ref{sec:specanalysis}. 
To improve the accuracy of the fits and due to the signal-to-noise ratios of our spectra, we require that the narrow components of all lines share the same width and redshift. 
In addition, we attempt to fit the \ion{Mg}{2}, \ion{C}{4}, \ion{He}{2}, \ion{C}{3}], and Balmer lines with a broad component.
We consider a spectrum to have a broad line if the fit with a broad line statistically improves the fit at the 95\% level as determined by an F-test.
In this work we define a broad line to have a minimum dispersion of 500\,km\,s$^{-1}$, i.e.\ a full width at half maximum (FWHM) $>1200$\,km\,s$^{-1}$ based on the delineation outlined in \citet{haoActiveGalacticNuclei2005}.

Example MMT spectra from this study are presented in Figure \ref{fig:examples} along with their corresponding \texttt{GELATO} fits.
The spectra are selected to be representative examples of the spectral classifications we use throughout this work in both redshift regimes.
\texttt{GELATO} proves to be an effective tool for recovering line fluxes, measuring equivalent widths, and detecting the presence of broad emission, enabling the spectral classification of our MMT follow-up spectroscopy.

We classify a galaxy as a Type I AGN if the spectrum shows evidence for at least one broad line and find a total of 99 Type I AGNs in our sample. 
We target 114 class I galaxies and secure 109 redshifts.
Of these, 93 are identified as Type I AGNs, 43 of which are in the low-redshift subsample and 48 in the high-redshift subsample.  
We target 63 class II galaxies and secure 30 redshifts.
Of these, six are identified as Type I AGNs, four of which are in the low-redshift subsample and two in the high-redshift subsample.
Broad emission lines for class I Type I AGNs display larger velocity widths with a median velocity FWHM of $\sim7800$\,km\,s$^{-1}$ and a standard deviation of $\sim2900$\,km\,s$^{-1}$ compared to a median FWHM of $\sim3400$\,km\,s$^{-1}$ and a standard deviation of $\sim900$\,km\,s$^{-1}$ for class II.

\begin{figure*}[ht!]
    \centering
    \includegraphics[width=\textwidth]{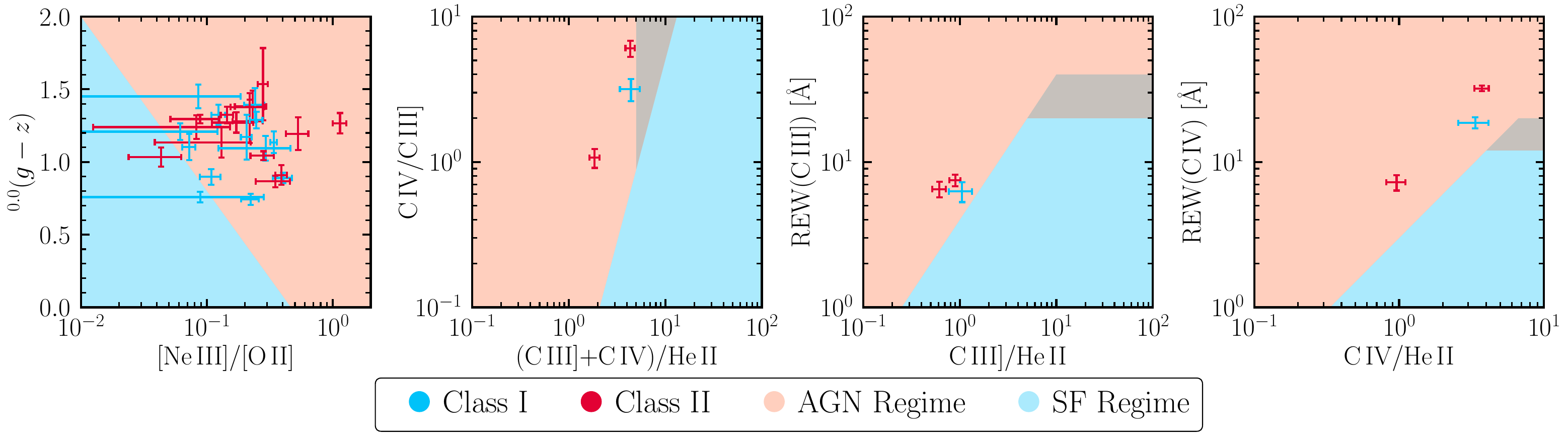}
    \caption{Emission-line diagrams for MMT follow-up targets with identified redshifts that show evidence for emission lines but do not exhibit broad emission or [\ion{Ne}{5}] emission. In the leftmost panel we present the TBT diagram for our low-redshift ($z < 1.2$) AGN candidates. In the three rightmost {panels} we present UV emission-line diagnostic diagrams from \citep{nakajimaVIMOSUltraDeep2018} for our high-redshift ($z \geq 1.2$) AGN candidates.\label{fig:eldiag}}
\end{figure*}

For the remaining 40 galaxies with a spectroscopic redshift and no detected broad line we attempt to classify the spectra based on the narrow emission lines using emission-line diagrams to distinguish ionization driven AGN activity from star formation (SF).
A common tool is the \citet[BPT;][]{baldwinClassificationParametersEmissionline1981} diagnostic, which uses ratios from common, strong optical emission lines to determine the source of the ionizing radiation.  
However, only one galaxy in our sample has spectral coverage of the necessary lines to be placed on the BPT diagram.
We therefore make use of the \citet[TBT;][]{trouilleOPTXProjectIdentifying2011} diagram which compares the [\ion{Ne}{3}] to [\ion{O}{2}] ratio to the rest-frame $g - z$ color and is more suited to the redshift range of our spectra ($z_\textrm{min} \simeq 0.23$).
We find that the results are consistent for the object which can be placed on both diagrams, finding that it lies on the border between a star-forming galaxy and an AGN.

To perform the K-correction to generate the rest-frame colors necessary for the TBT diagnostic, we fit a combination of the elliptical, spiral, and irregular empirical galaxy templates from \citet{assefLowResolutionSpectralTemplates2010} combined with {the normal, as opposed to warm- or hot-dust deficient, AGN template from  \citet{lyuIntrinsicFarinfraredContinua2017} and AGN extinction curve from Lyu (in prep.)\footnote{{See Section 2.2 in \citet{lyuInfraredSpectralEnergy2022} for details}.}}
to the HSC and WISE photometry at the spectroscopic redshift of the galaxy using the procedure described in Section 4.1 of \citet{hvidingNewInfraredCriterion2022}. 
The rest-frame colors and associated errors are computed by convolving the SDSS $g$ and $z$ filters \citep{gunnSloanDigitalSky1998} with the full set of models from the MCMC chain shifted into the rest frame.
The leftmost panel of Figure \ref{fig:eldiag} presents the TBT diagram for the objects from our low-redshift subsample that do not have evidence for a broad line or [\ion{Ne}{5}], 14 from class I and 16 from class II.

For the high-redshift subsample without detected broad lines, we make use of the diagnostics presented in \citet[N18;][]{nakajimaVIMOSUltraDeep2018} which compare the ratios and rest-equivalent widths (REWs) of the \ion{C}{4}, \ion{He}{2}, and \ion{C}{3}] emission lines. 
We present the three high-redshift, narrow-line galaxies on the N18 diagrams in the {rightmost} panels of Figure \ref{fig:eldiag}.
For galaxies placed on either the TBT or the N18 diagrams we sample the posterior to compute the probability that each galaxy is in the AGN regime.
Conversely, the complementary probability can be calculated that the emission is not consistent with AGN activity and instead due to SF.

In addition we investigate which galaxies show evidence for the coronal emission line [\ion{Ne}{5}] that is a reliable indicator of AGN activity due to its high ionization potential of $\sim$100\,eV \citep[and references therein]{mignoliObscuredAGNZCOSMOSBright2013,feltreNuclearActivityStar2016,negusCatalog71Coronal2023}.
We find nineteen narrow-line galaxies, seven from class I and twelve from class II, with robust [\ion{Ne}{5}] detections (flux SNR $>$ 3), five of which have detections at the 10$\sigma$ level,  one from class I and four from class II. 
We note that 17 of the [\ion{Ne}{5}]-detected galaxies are already classified as AGN using the emission line diagrams at the 3$\sigma$ confidence level, with the last two classified at the $\sim80\%$ and $\sim90\%$ level.

To place constraints on the non-AGN contamination fraction in classes I and II, we assign each object an AGN probability. 
Type I and Type II [\ion{Ne}{5}]-detected AGNs are assigned a probability of 100\%.
For the remaining narrow-line galaxies we compute the AGN probability by sampling the posterior of each source one million times and calculating the fraction that lie in the AGN regime.
As there are three N18 diagrams, we take the mean AGN probability for each of the three narrow-line high-redshift galaxies but find that the result is consistent regardless of which diagnostic is used.

Two class II galaxies in our sample show little-to-no evidence for emission lines used in AGN diagnostics. 
These objects cannot be robustly classified as AGNs nor as SF galaxies. 
To place lower bounds on the AGN fractions for our classes, we treat these objects as having a 0\% probability of being an AGN and label them as Other in the forthcoming analysis. 

\begin{deluxetable}{c|r|r|r|r|r}
\label{tab:fracs}
\tablecaption{AGN Fraction by Class}
\tablehead{\colhead{Class} & \colhead{Type I} & \colhead{Type II} & \colhead{SF} & \colhead{Other} & \colhead{No-$z$}}
\startdata
\multicolumn{6}{c}{Targets with Redshift}\\
\hline
I & 86.24\% & 12.31\% & 1.45\% & 0.0\% & \multicolumn{1}{c}{---} \\
II & 20.0\% & 65.38\% & 7.95\% & 6.67\% & \multicolumn{1}{c}{---} \\
\hline
\multicolumn{6}{c}{All Targets}\\
\hline
I & 82.46\% & 11.77\% & 1.39\% & 0.0\% & 4.39\% \\
II & 9.52\% & 31.13\% & 3.79\% & 3.17\% & 52.38\%
\enddata
\end{deluxetable}

\begin{figure}[ht!]
    \centering
    \includegraphics[width=\columnwidth]{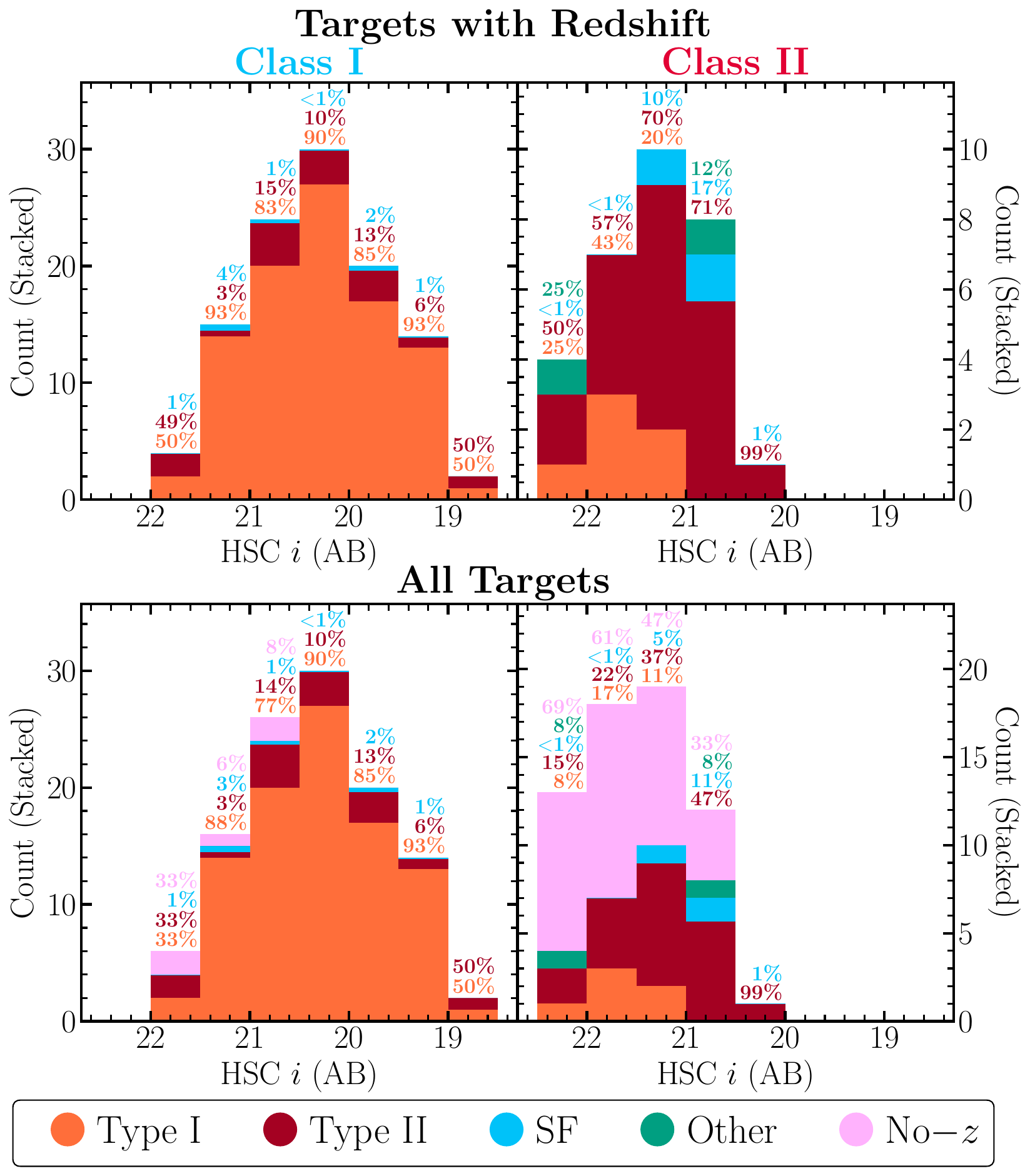}
    \caption{Spectroscopic classification as a function of optical magnitude for class I (left) and II (right). 
    The top panels present the fractions for targets with a secure redshift, while the bottom panels show the percentages accounting for objects without an identified redshift.
    Class I is dominated by Type I AGNs with minimal contamination from non-AGN.
    The majority of class II are Type II AGNs with a higher fraction of non-AGN contaminants.
    Overall fractions for each class are presented in Table \ref{tab:fracs}.
    \label{fig:agnfrac}}
\end{figure}

By taking into account the probability that each source is an AGN, we compute the average breakdown of classes I and II by spectroscopic classification and present the results in Table \ref{tab:fracs}.
Figure \ref{fig:agnfrac} presents the spectroscopic classification as a function of optical magnitude and class with and without taking into account those objects without identified redshifts.
Class I is dominated by Type I AGNs, with a $<$3\% contamination rate, while the majority of class II is mostly comprised of Type II AGNs with a moderate contamination rate of $\sim$15\% from non-AGN sources.
We note that our redshift recovery fraction is relatively low at the faintest magnitudes for class II, with over half of the class without an identified redshift.
We therefore present the lower-limit AGN fractions as well, where only half of class II are identified as AGN with a confident redshift.
While the spectroscopic classification of class I has little-to-no evolution with optical magnitude, class II has a higher fraction of Type I AGNs for fainter targets. 
This is likely driven by a bias where broad-line AGNs are easier to identify spectroscopically than Type II AGNs at the same brightness due to their strong emission, though deeper spectroscopy in class II would be needed to confirm this. 
Overall, our ML-selected classifications are effective at recovering AGN samples with a high level of accuracy. %, which we discuss further in Section \ref{sec:conc_fut}.

\begin{figure*}[ht!]
    \centering
    \includegraphics[width=0.75\textwidth]{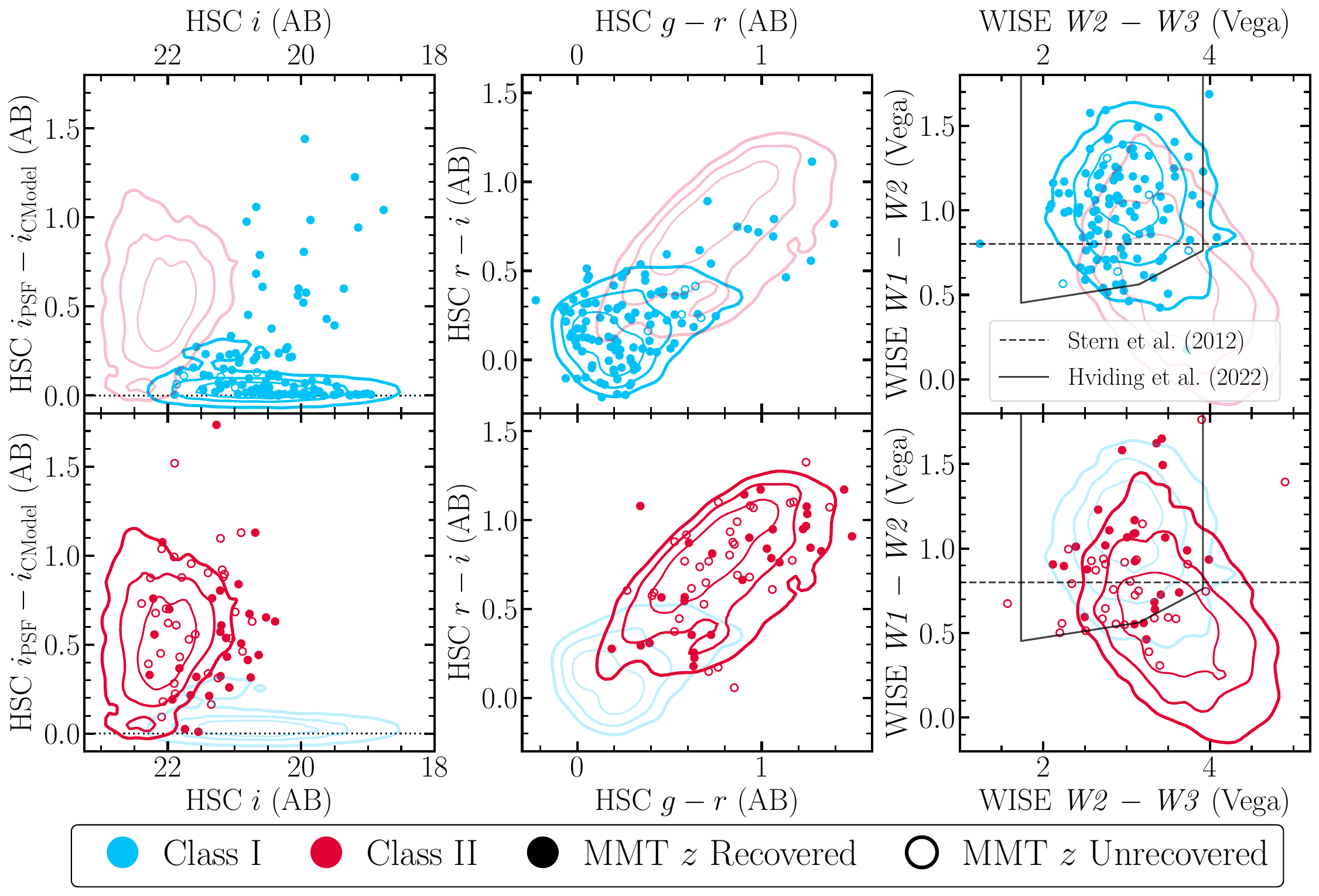}
    \caption{Optical magnitude-morphology (left), optical color-color (center), and mid-IR color-color (right) for classes I (blue; top) and II (red; bottom).
    We plot contours encompassing 25\%, 50\%, and 75\% of the classes and plot our MMT follow-up targets with recovered and unrecovered redshifts as {filled and unfilled circles} respectively.
    Mid-IR color-color contours are restricted to objects with SNR $>3$ in \textit{W1}, \textit{W2}, and \textit{W3}.
    We plot the \citet[solid]{hvidingNewInfraredCriterion2022} and \citet[dashed]{sternMidinfraredSelectionActive2012} mid-IR selections to highlight regions where typical mid-IR selected AGNs inhabit.
    We note that our follow-up spectroscopy span the majority of each class's parameter space, but are limited to $i < 22.5$, which particularly effects class II.\label{fig:color}}
\end{figure*}

\begin{figure*}[ht!]
    \centering
    \includegraphics[width=\textwidth]{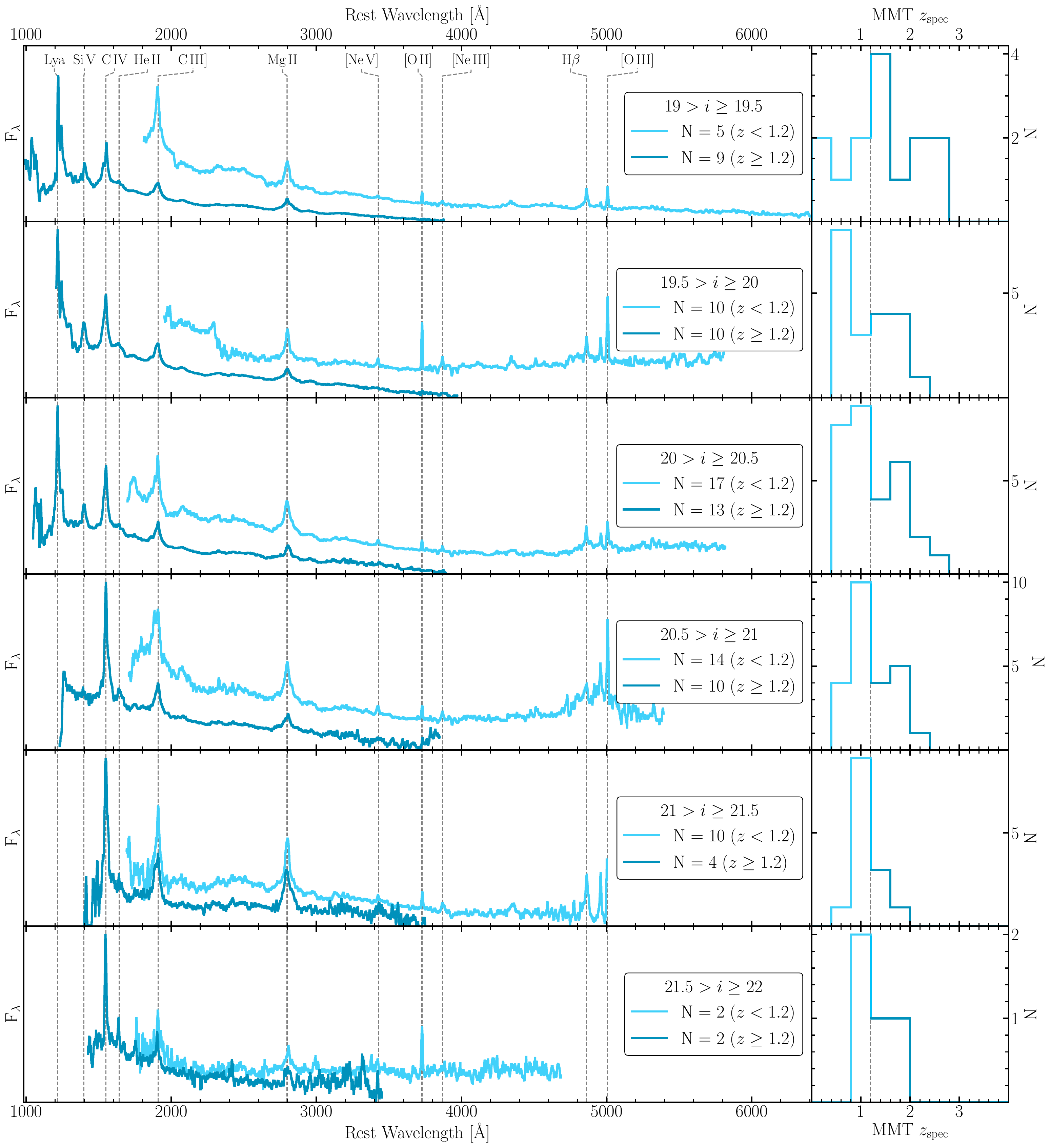}
    \caption{Low-redshift ($z < 1.2$; light blue) and high-redshift ($z \geq 1.2$; dark blue) stacked MMT spectra (left) and redshift distributions (right) for class I by $i$-band magnitude. The class is primarily comprised of Type I broad-line AGNs. Stacked spectra are smoothed with a boxcar function of width of 10\AA\ for plotting purposes only.\label{fig:stackingI}}
\end{figure*}

\begin{figure*}[ht!]
    \centering
    \includegraphics[width=\textwidth]{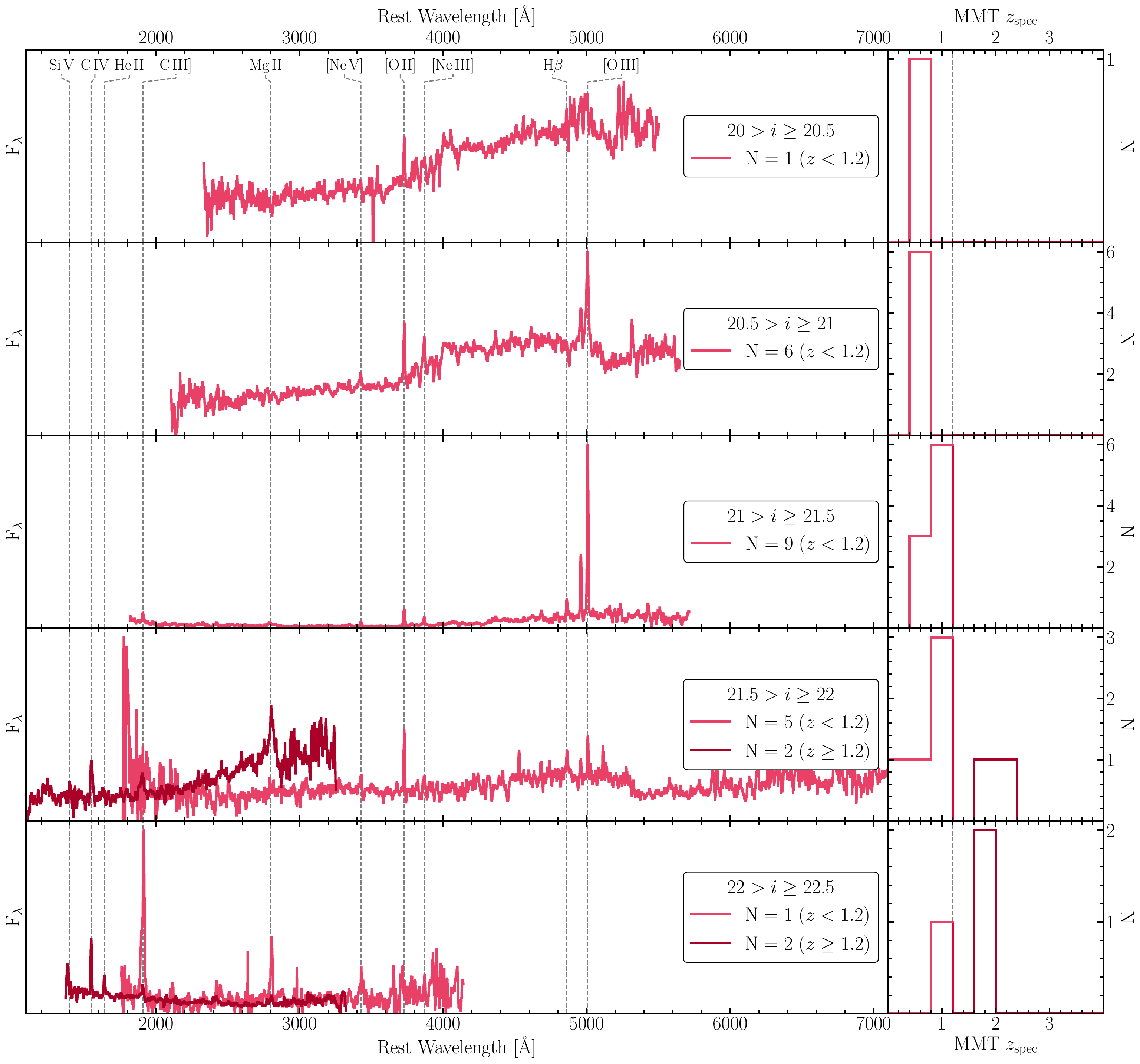}
    \caption{Low-redshift ($z < 1.2$; light red) and high-redshift ($z \geq 1.2$; dark red) stacked MMT spectra (left) and redshift distributions (right) for class II by $i$-band magnitude. The class is primarily comprised of Type II narrow-line AGNs. Stacked spectra are smoothed with a boxcar function of width of 10\AA\ for plotting purposes only.\label{fig:stackingII}}
\end{figure*}

MMT spectroscopy has confirmed that our ML-selected AGN classes are comprised of AGNs. 
Class I selects $\sim$20\,deg$^{-2}$ more Type I quasars as compared to SDSS quasar selected samples. 
These additional objects are primarily composed of galaxies with redder optical colors than the SDSS selection.
Conversely class II predominantly selects Type II quasars with an on-sky density of $\sim$200\,AGNs\,deg$^{-2}$.
Even when taking the lower limit when considering objects without a recovered redshift, class II represent an on-sky density of $\sim$100\,deg$^{-2}$ predominantly Type II AGNs.
The few matched SDSS spectra that exist for class II primarily come from emission-line galaxy subsurveys and are limited to the brightest and bluest subset of the class.

\subsection{Spectral Stacking}\label{subsec:stack}

To determine if the AGN properties are consistent across our ML-selected classes, we investigate our spectroscopy in bins of magnitude, redshift, and optical color and compare to the spectral stacks of matched SDSS galaxies.
In Figure \ref{fig:color} we present the optical magnitude-morphology, optical color-color and mid-IR color-color distributions of our selected targets with respect to our parent HSC sample.
Our follow-up targets are representative across the parent classes' morphological and optical color-color spaces.
We therefore stack the spectra of AGNs in our follow-up to accurately measure the properties of classes I and II. 
We consider an object an AGN if it has an AGN probability greater than 50\% as determined in Section \ref{subsec:emiss}.
By combining spectra of similar objects, we can enhance the SNR of specific spectral features, such as emission lines, and study their properties as a function ML-selected class.

Spectra are stacked in the low-redshift and high-redshift subsamples to determine what impact, if any, redshift has on the optical colors of the classes and to ensure all spectra in the subsample share a region of continuum free from spectral features.
We normalize all spectra to a region of the continuum: 3500-3600\AA\ for the low-redshift subsample and 2000-2100\AA\ for the high-redshift subsample.
All spectra are interpolated onto a wavelength grid of 1\AA\ intervals using the \texttt{SpectRes} \citep{carnallSpectResFastSpectral2017,carnallSpectResSimpleSpectral2021} flux-preserving spectral resampling code.
The weighted arithmetic mean is taken at each wavelength to generate the stacked spectrum.

\subsubsection{Stacks by Optical Magnitude}\label{subsubsec:stackmag}

In Figure \ref{fig:stackingI} we present the stacks of class I AGNs in bins of optical magnitude along with the distributions of the retrieved spectroscopic redshift.
We observe that class I is dominated by Type I AGNs with broad optical and UV emission lines and strong blue optical/UV continua. 
This trend continues across the entire optical magnitude range of the sample.
In addition, we find no difference of the spectroscopic properties between the high-redshift and low-redshift subsamples.

We repeat our stacking procedure for those sources in class II.
Figure \ref{fig:stackingII} presents the stacks of class II AGNs in bins of optical magnitude along with the distributions of the retrieved spectroscopic redshift.
We observe that class II is predominantly comprised of Type II AGNs with strong narrow emission lines and clear stellar continuum as evidenced by stellar absorption features.
This trend continues across the entire optical magnitude range of the sample but with a variety in the emission-line strength of the stacks. 
In addition, while the properties of the low- and high-redshift samples of class II appear consistent, it is difficult to confirm with relatively few high-redshift class II objects. 

\subsubsection{Comparison to SDSS Spectral Stacks} \label{subsubsec:stacksdss}

While SDSS-matched class I galaxies are nearly all ($>99$\%) identified as QSOs by the automated SDSS pipeline, this is only true for 22\% of SDSS-matched class II galaxies. 
This does not mean that the remaining 79\% do not host AGN, but rather that there are no detectable broad-lines in the spectrum or that they suffer from greater contamination from the host.
We thus create spectral stacks for class II galaxies with existing SDSS spectroscopy.
Figure \ref{fig:sdssstack} presents these stacks compared to stacks of our class II MMT spectra. 
While it is immediately apparent that the MMT follow-up presented in this work samples a redder distribution of galaxies than the SDSS-matched sample, it is also clear that the SDSS subsample hosts a high fraction of AGN at both high redshift, as evidenced by broad emission features, and at low redshift, as evidenced by the presence of [\ion{Ne}{5}] coronal emission and high-ionization emission line ratios from [\ion{O}{3}]/H$\beta$ and [\ion{Ne}{3}]/[\ion{O}{2}].
Therefore, the SDSS-matched class II galaxies are indeed comprised of Type II AGN, similar to the bulk of class II, but are bluer than the majority of the parent class.

\subsubsection{Stacks by Optical Color}\label{subsubsec:stackcolor}

\begin{figure*}[ht!]
    \centering
    \includegraphics[width=\textwidth]{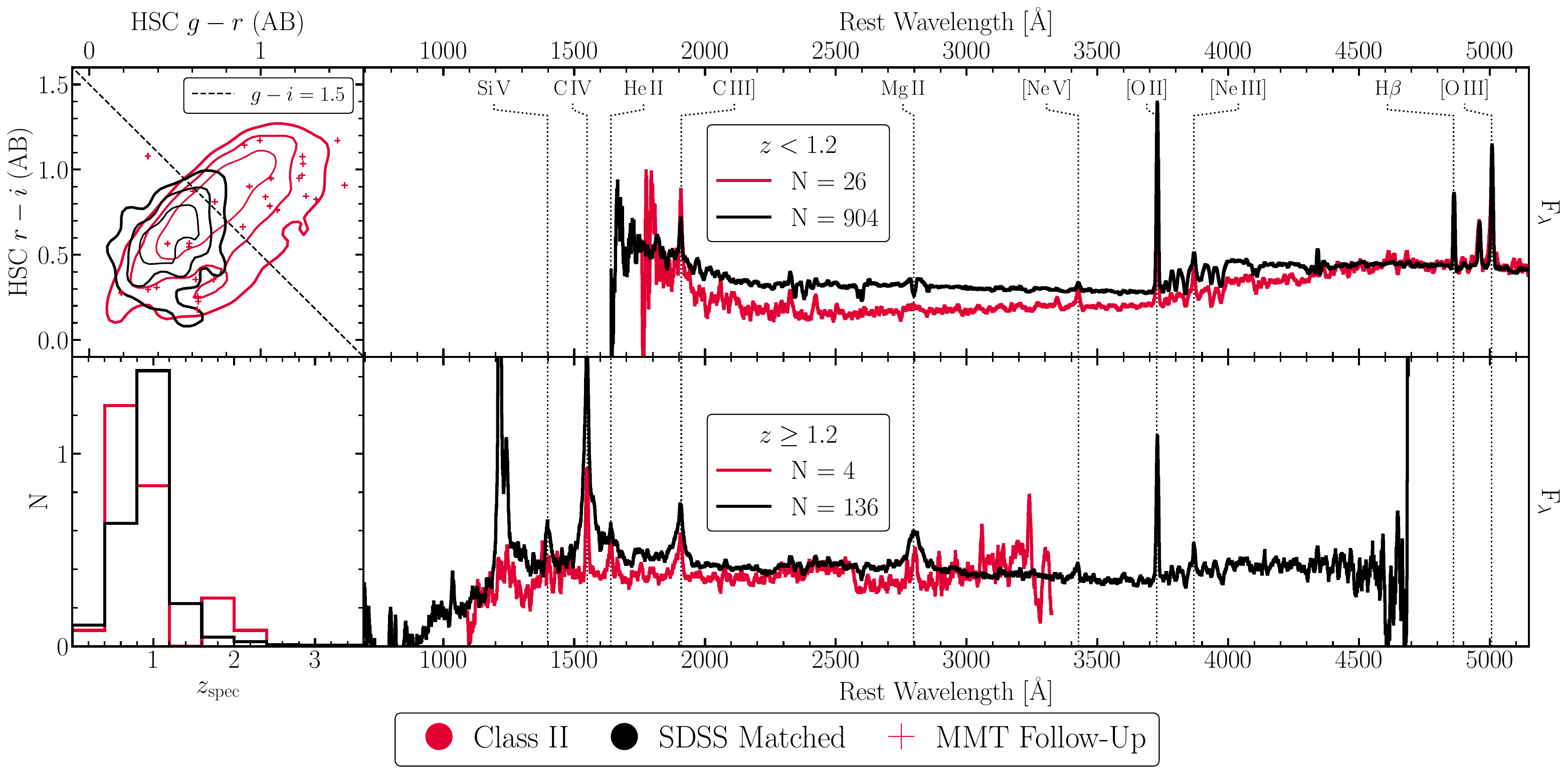}
    \caption{Optical color-color (top left) and spectroscopic redshift (bottom left) distributions of class II (red) and SDSS matched (black) subsamples.
    In addition, we present stacked spectra for low-redshift ($z < 1.2$; top right) and high-redshift ($z \geq 1.2$; bottom right) subsamples.
    Stacked spectra are smoothed with a boxcar function of width of 10\AA\ for plotting purposes only.
    \label{fig:sdssstack}}
\end{figure*}

A clear advantage of our ML-identified classes is the identification of redder AGN candidates than typical spectroscopic surveys as highlighted in Section \ref{subsec:sample} and \ref{fig:sdssstack}.
In particular, it is of interest to determine the AGN properties of the candidates that are redder than the matched SDSS spectroscopy, to determine the differences to their bluer counterparts. 
While our stacking analysis has confirmed that classes I and II are indeed comprised of AGNs across all magnitudes, we pursue stacking in bins of optical color to determine the difference in AGN properties, if any, between the bluest and reddest AGNs in our sample.

\begin{figure*}[ht!]
    \centering
    \includegraphics[width=\textwidth]{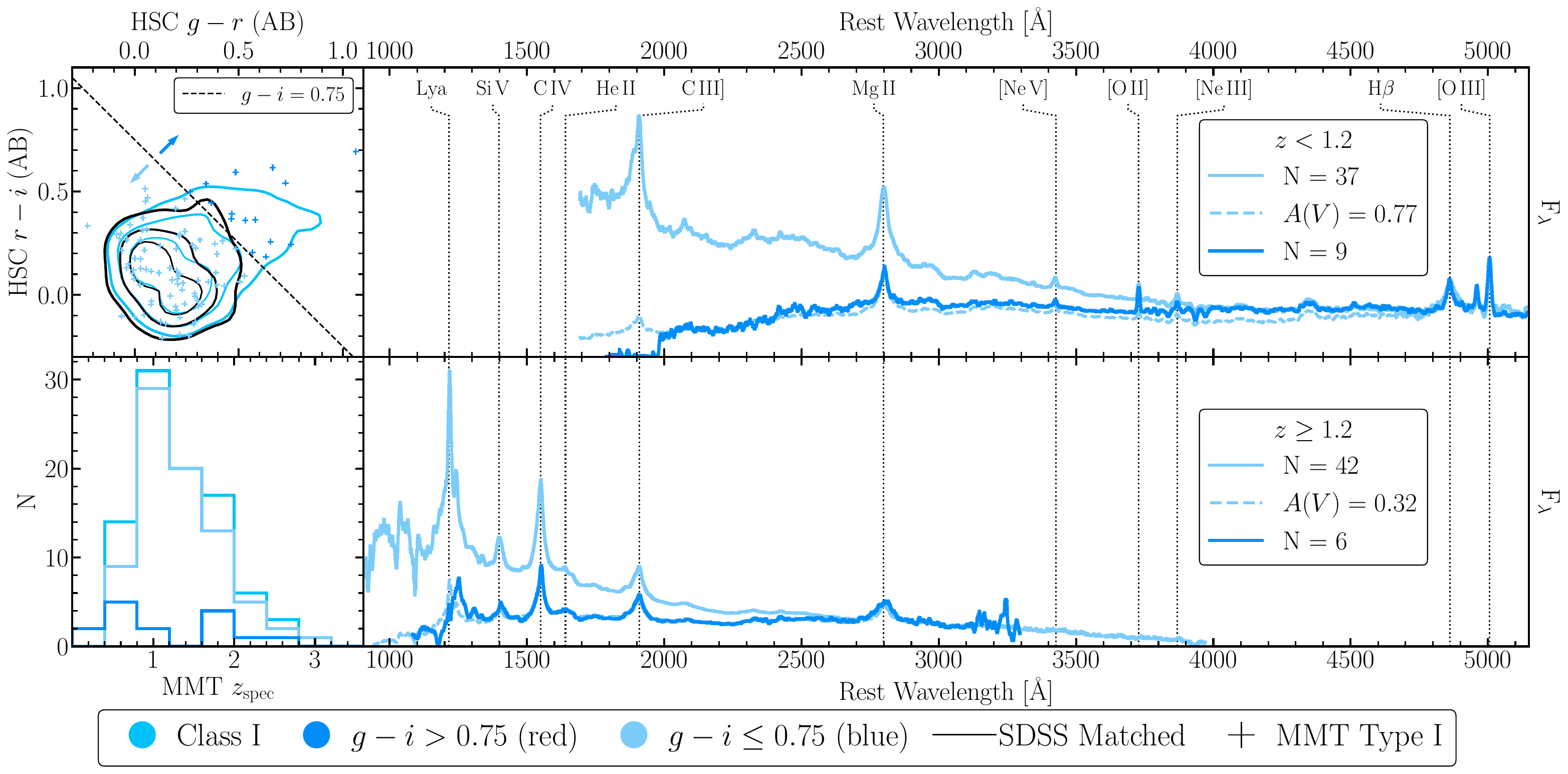}
    \caption{Optical color-color (top left) and MMT redshift (bottom left) distributions of class I (blue) and ``red'' ($g-i>0.75$; dark blue) and ``blue'' ($g-i\leq0.75$; light blue) subsamples.
    In addition, we present stacked spectra for low-redshift ($z < 1.2$; top right) and high-redshift ($z \geq 1.2$; bottom right) subsamples for red and blue class I targets identified as Type I AGNs.
    We fit the red stacked spectra with the blue stacked spectra combined with the SMC Bar extinction curve. Stacked spectra are smoothed with a boxcar function of width of 10\AA\ for plotting purposes only.
    \label{fig:colorstackI}}
\end{figure*}

To investigate the properties of our AGNs in class I as a function of optical color, we split the class into a ``red'' subsample ($g-i>0.75$) and a ``blue'' subsample ($g-i\leq0.75)$ based on where we observe class I deviates from SDSS-matched spectroscopy in Figure \ref{fig:sdss}.
In addition, we restrict our analysis to Type I AGNs, the dominant constituent (85\%) of class I.
Figure \ref{fig:colorstackI} presents the stacks as a function of redshift and optical magnitude.
Regardless of optical color or redshift, the emission-line shape and strength are consistent, suggesting Type I AGNs in class I are drawn from a similar underlying population of galaxies.

To determine if the differences between the ``red'' and ``blue'' subsamples can be explained solely through obscuration, we fit the red stacked spectrum by taking the the blue stacked spectrum and applying an extinction curve.
We make use of the Small Magellanic Cloud (SMC) Bar extinction curve \citep[$R(V)=2.74$]{gordonQuantitativeComparisonSmall2003} as it has been found to best describe the extinction in dust-reddened quasars and AGNs \citep{richardsRedReddenedQuasars2003,hopkinsDustReddeningSloan2004}.
By minimizing the $\chi^2$ we retrieve a best fit extinction of $A(V)=0.77$\,mag and 0.33\,mag for the low- and high-redshift $g-i > 0.75$ subsamples respectively. 
The extincted blue stacked spectra reproduces the red stacked spectra, suggesting the diversity in Type I class I objects is driven by extinction, potentially nuclear or galaxy-wide. 
Our ML selection methodology is therefore able to recover AGNs at higher obscuration levels.
Although reddening can naturally explain the spectral shapes of the red tail of Class I objects, we cannot entirely rule out the contribution of a redder stellar continuum for these sources.

\begin{figure*}[ht!]
    \centering
    \includegraphics[width=\textwidth]{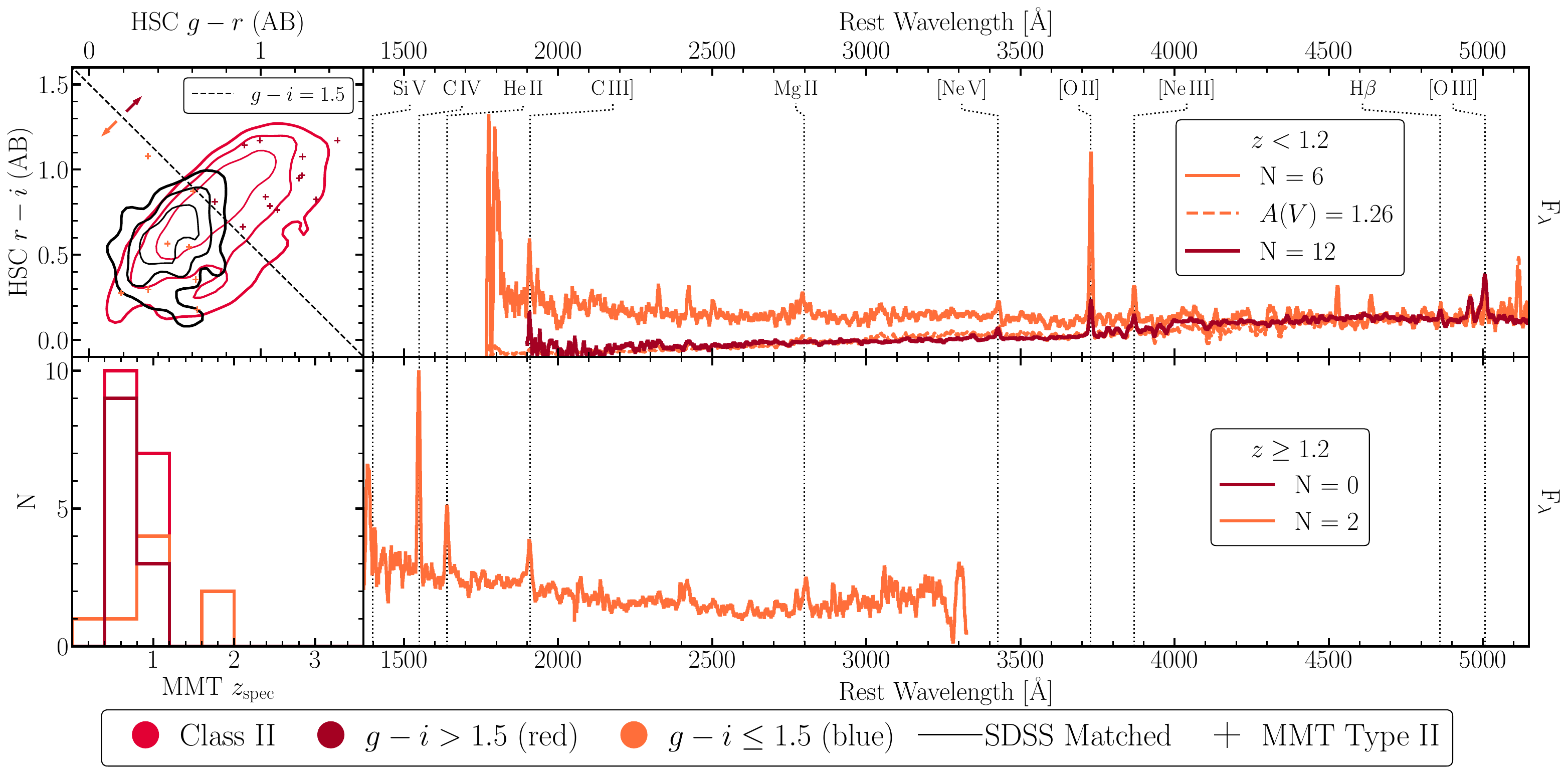}
    \caption{Optical color-color (top left) and MMT redshift (bottom left) distributions of class II (red) and ``red'' ($g-i>1.5$; dark red) and ``blue'' ($g-i\leq1.5$; orange) subsamples.
    In addition, we present stacked spectra for low-redshift ($z < 1.2$; top right) and high-redshift ($z \geq 1.2$; bottom right) subsamples for red and blue class II targets identified as Type II AGNs.
    We fit the 'red' stacked spectra with the 'blue' stacked spectra combined with the SMC Bar extinction curve. Stacked spectra are smoothed with a boxcar function of width of 10\AA\ for plotting purposes only.
    \label{fig:colorstackII}}
\end{figure*}

To investigate the properties of our class II follow-up spectroscopy as a function of optical color, we split the class based on the $g-i$ color.
Since class II is optically redder than class I, we define the ``red'' subsample to have $g-i>1.5$, and the ``blue'' subsample as $g-i\leq1.5$.
In addition, we restrict our analysis to Type II AGN, the majority constituent (65\%) of class II. 
Figure \ref{fig:colorstackII} presents the stacks as a function of redshift and optical magnitude.
Similar to class I, the emission-line shape and strength are consistent across optical magnitude, although we are unable to draw any strong conclusions across redshift as we only have two blue high-redshift Type II class II AGNs, and no corresponding red galaxies.
We again fit the red stacked spectra with the blue stacked spectra by applying the SMC bar extinction curve, however this can only be done at low-redshift as there are no `red', high-redshift, spectra in our sample. 
We find a best-fit $A(V)=1.26$\,mag, suggesting Type II class II AGNs are distinguished primarily by varying levels of extinction, but deeper spectroscopy would be required to constrain the differences, if any, in the underlying stellar continua.

\subsection{Balmer Decrement} \label{subsec:baldec}

To further investigate the level of extinction in the ML-selected classes, we measure the Balmer decrement in our spectroscopic sample.
Typically the H$\alpha$/H$\beta$ line ratio is used as the Balmer decrement however the redshift of our galaxies combined with our spectral coverage requires that we measure the H$\gamma$/H$\beta$ line ratio to compute the effects of attenuation.
While measurements of $A(V)$ from H$\gamma$/H$\beta$ are not as robust as those measured from the traditional H$\alpha$/H$\beta$ ratio, we aim to bolster our interpretation that AGN in our sample exhibit obscuration, rather than precise measurements of extinction in specific galaxies.
In this work we assume case B recombination in which the value of the H$\gamma$/H$\beta$ ratio is 0.47 \citep{osterbrockAstrophysicsGaseousNebulae2006}.

\begin{figure}[ht!]
    \centering
    \includegraphics[width=\columnwidth]{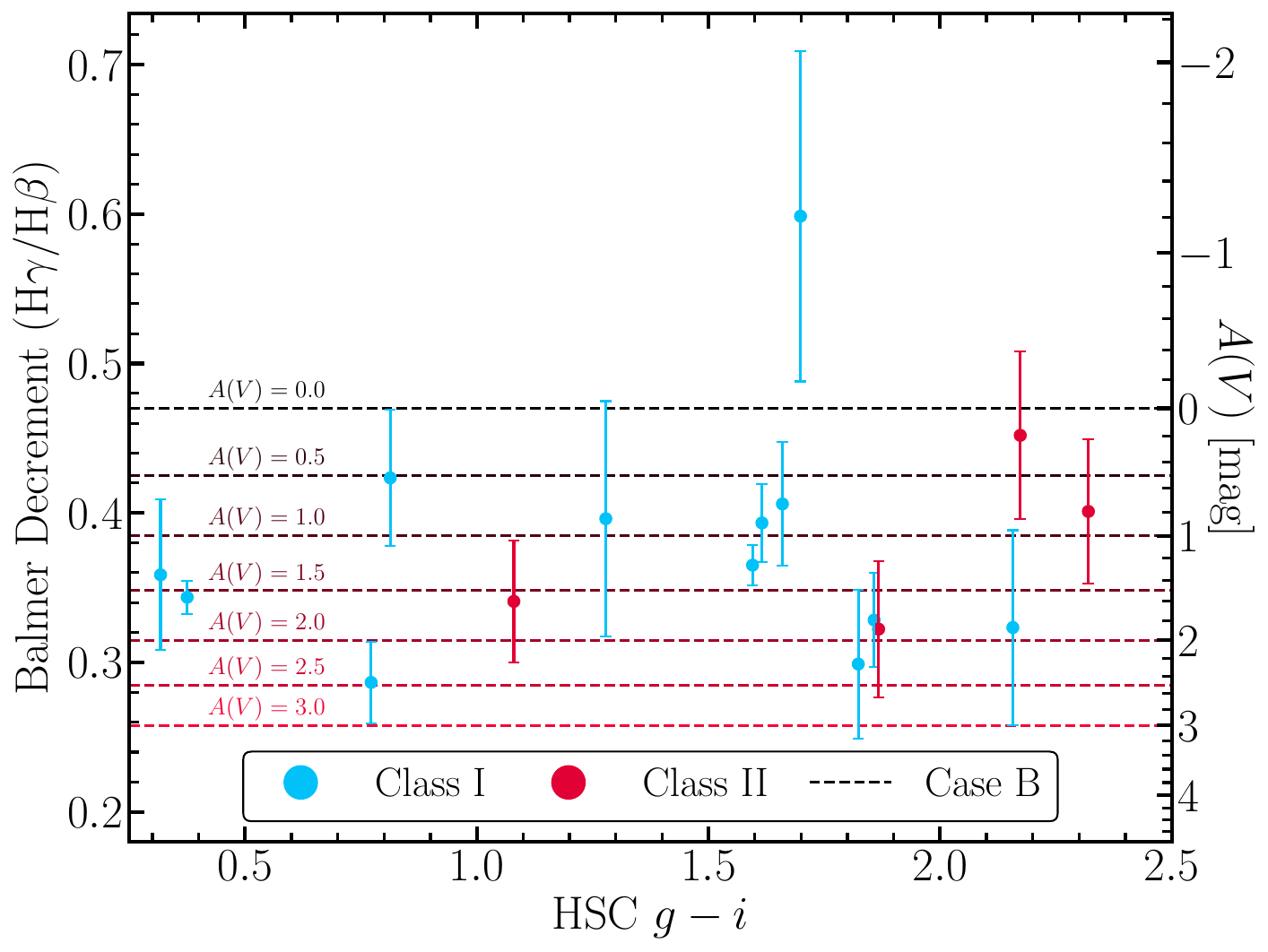}
    \caption{Balmer decrement (H$\gamma$/H$\beta$ (left) and attenuation (right) versus $g-i$ color for class I (blue) and class II (red) objects with H$\beta$ and H$\gamma$ flux SNRs $>$ 5.
    We plot horizontal lines corresponding the the necessary $A(V)$ required to achieve a specific Balmer decrement assuming the SMC Bar extinction curve.
    \label{fig:baldec}}
\end{figure}

We plot the Balmer decrement for spectra with confidently measured H$\beta$ and H$\gamma$ narrow-line fluxes (SNR $>$ 5) against spectroscopic redshift in Figure \ref{fig:baldec}.
Using the SMC Bar extinction curve, we compute the requisite level of attenuation implied by the measured Balmer decrement.
While we can only obtain the requisite SNR for 16 objects, 11 from class I and five from class II, the spectra exhibit attenuations ranging from $A(V)\simeq0 - 2.5$\,mag, further indicating that extinction is what drives the diversity in the optical colors of AGNs within classes I and II.

As we investigate the narrow-line Balmer decrement, our measurements are sensitive to extinction on larger scales than the nucleus.
We measure a weighted-mean attenuation of $A(V) \simeq 1.3$ for both class I and class II, representative of significant galaxy-scale obscuration for these sources.
However, due to the few galaxies for which we can perform this analysis and the larger uncertainties on the recovered lines, especially for the weaker H$\gamma$, further investigation would be required to draw conclusions about each class.
This is especially true for class I, which is comprised nearly entirely of Type I AGN, as our analysis relies on the decomposition of broad and narrow emission, which can contribute an additional source of uncertainty. 

\section{Discussion}
\label{sec:disc}

Our ML-selected classes represent substantial AGN populations over the entire $\sim$1,000\,deg$^2$ HSC footprint. 
The spectroscopic analysis conducted in this work demonstrates that the high number densities of class I and II are not driven by contamination and are indeed comprised of accreting SMBHs.
Accounting for the AGN-selection accuracy, the on-sky density of class I equates to $\sim$145\,AGNs\,$\deg^{-2}$ while class II represents over $\sim$200\,AGNs\,$\deg^{-2}$.
Even when taking into account targets for which we could not retrieve a redshift, which primarily affects class II, the classes together make up $\gtrsim$240\,AGNs\,$\deg^{-2}$.
Classes I and II therefore represent some of the densest AGN samples over large areas, especially compared to samples selected from SDSS or WISE.

Critically, a substantial fraction of these galaxies lie below the detection limits of typical WISE criteria or inhabit in regions of mid-IR and optical color space that typically suffer heavy contamination from star-forming galaxies.
This is especially true at the reddest optical colors and faintest optical magnitude, where there is limited existing SDSS spectroscopy that recovers these AGN populations. 
Given that our spectroscopic analysis suggests that the red optical colors in our samples are driven primarily by dust extinction, rather than stellar properties or redshift, our ML-selected AGNs represent a window into low-luminosity and/or obscured SMBH growth for large samples.

In particular, the reddest class I Type I AGNs appear to exhibit similar spectral properties to the reddest quasars studied in \citet{glikmanFIRST2MASSRedQuasar2007,glikmanWISE2MASSSurveyRed2022,banerjiHeavilyReddenedType2015,hamannExtremelyRedQuasars2017}. 
\citet{fawcettFundamentalDifferencesProperties2022} find that red quasars are distinguished primarily from `blue quasars' through extinction with measured values of $A(V)$ reaching 0.7\,mag, consistent with the extinction of our low-redshift red spectral stack.
A similar trend with extinction was additionally observed for Type I quasars in \citep{patReconstructingClassifyingSDSS2022}, also through the use of unsupervised ML.
Our ML approach may present an effective methodology to finding red quasars along with blue quasars at higher levels of AGN-selection accuracy than traditional spectroscopic quasar searches. 

Class II is predominantly comprised of Type II AGNs.
Historically, optical spectroscopy has been an effective tool for assembling samples of Type II AGNs, though their demographics are not fully understood \citep{reyesSpaceDensityOptically2008,yuanSpectroscopicIdentificationType2016}.
Class II appears to recover Type II AGNs similar to previous study and may additionally encompass AGNs similar to those selected from small-area X-ray surveys \citep{lussoBolometricOutputHostgalaxy2011} or to the Type II AGNs identified in \citet{hvidingNewInfraredCriterion2022} which cannot be selected using mid-IR colors alone. 
Our selection suggests these sources can be directly targeted using a combination of optical imaging and mid-IR data without resorting to X-ray observations or subsets of emission line galaxy samples.

Critically, these galaxies likely represent an opportunity to understand weaker AGN activity relative to the host galaxy and at higher redshifts. 
As simulations suggest a galaxy's SMBH spends a larger fraction of its life as a low-Eddington-ratio AGNs rather than a poweful quasar, it remains a priority to fully detect and characterize the low-luminosity AGN population \citep{novakFeedbackCentralBlack2011,schawinskiActiveGalacticNuclei2015}.
Class II AGNs may reflect the impact of low-luminostiy AGN feedback on the host galaxy as they co-evolve through cosmic time, or even for testing the accuracy of model predictions for the time-scales of low-Eddington-rate SMBH accretion. 

The combination of optical imaging with mid-IR photometry presents a powerful AGN selection methodology. 
Mid-IR colors, sensitive to obscured and unobscured SMBH accretion, paired with optical colors, sensitive to host-galaxy properties, enables the retrieval of accurate AGN samples.
This is especially true when paired with additional discerning information, such as the optical morphologies used in this work, to provide additional insight into the galaxy's dynamical history.
% The Quaia catalog, for example, selects $\sim$1,300,000 quasars through the combination of Gaia optical photometry and unWISE colors \citep{storey-fisherQuaiaGaiaunWISEQuasar2023}. 
% While they don't characterize the morphological shape of their sources, the Quaia sample critically folds in proper motion data to drastically reduce their fraction of interlopers. 
The inclusion of additional, non-color, information, such as morphology, proves to be an effective approach for selecting accurate samples of AGNs across the entire sky that could not previously be recovered solely by using photometric colors from one wavelength regime. 

\section{Conclusions \& Future Work}\label{sec:conc_fut}

In this work we perform optical spectroscopic follow up of AGN candidates selected from a parent sample of WISE-matched HSC targets using ML.
The optical morphological data paired with the longer wavelengths afforded by WISE present a powerful dataset for selecting AGN candidates which can be fully explored through our use of ML algorithms.
We present the results of our MMT spectroscopy which enable the characterization of our ML-selected classes, specifically with respect to AGN type and selection accuracy.

We recover redshifts in $\sim$80\% of our 178 MMT spectra drawn from our target classes.
Our spectroscopy confirms that classes I and II are predominately comprised of AGNs, reiterating that deep optical imaging joined with mid-infrared data is effective at selecting AGN candidates from photometry alone.
Class I has a small contamination rate ($<3\%$) and is dominated by Type I AGNs ($>90\%$) while class II has a moderate contamination rate ($\sim$15\%) and is mostly made up of Type II AGNs ($\sim$65\%).
The AGN-selection accuracy of our follow-up spectroscopy highlights the efficacy of deep optical imaging combined with mid-IR photometry especially when paired with ML techniques to analyze the multi-dimensional color-morphology parameter space. 

Critically, our AGN classes occupy an area of the parameter space that has not been explored by traditional spectroscopic surveys such as SDSS.
Our selected AGNs live in galaxies that are more diffuse and exhibit redder colors than the traditional spectroscopic samples.
In addition, we demonstrate that the difference between the ``red'' and ``blue'' stacked spectra can be explained by applying an extinction curve, suggesting that AGNs from the ML-selected classes are drawn from a continuum in extinction.
This is further supported by our investigation of the Balmer decrements in these sources that exhibit a range of attenuations from $A(V) \simeq 0$\,mag to $A(V) \simeq 2.5$\,mag.
Together this suggests our selection probes more obscured sources with a more easily detectable host galaxy, i.e.\ low-luminosity AGN activity.

In the future, we aim to extend our spectroscopy to fainter optical magnitudes.
This is especially relevant for class II, where we were restricted to the brightest 75\% of objects in this work, and primarily retrieved redshifts for the brightest 20\% of the class.
Class II follow-up would be ideal from MMT Binospec, a multiplexed high-throughput spectrograph that can probe fainter targets than MMT Hectospec, albeit over a smaller area \citep{fabricantBinospecWidefieldImaging2019}.
Additional spectroscopy in class II will also enable the individual characterization of the parent UMAP subclasses $b$, $c$, and $d$.

{The next generation of spectroscopic surveys from the Dark Energy Spectroscopic Instrument \citep[DESI;][]{desicollaborationEarlyDataRelease2023} or the upcoming 4-metre Multi-Object Spectrograph Telescope \citep[4MOST;][]{dejong4MOSTProjectOverview2019} will further inform the properties of ML-selected classes with spectra of fainter and redder galaxies/AGN.
Furthermore, ML selection techniques may also prove useful for mitigating quasar misidentification in spectroscopic validation pipelines which can impact the AGN recovery fraction by a few percent as seen in the SDSS and DESI surveys \citep{farrOptimalStrategiesIdentifying2020,alexanderDESISurveyValidation2023}.}
In addition, studying the X-ray and radio properties of our ML-identified AGN candidates may provide additional insight into the accretion properties of our sample and consequently their context relative to other AGN samples.

Finally, once the next generation of optical and near-IR survey telescopes come online, the ML methodology can be applied to select orders of magnitude more AGN candidates.
Rubin, for example, presents a similar filter set and depth to HSC, but whose Legacy Survey of Space and Time (LSST) will cover an area of the sky that is nearly an order of magnitude larger. 
The recovery of these new AGN populations will inform our understanding of galaxy and SMBH coevolution and provide follow-up targets for further in-depth study. 
The unique composition of optical and mid-IR data paired with ML techniques proves to be a powerful tool for selecting AGN candidates from the next generation of surveys that are currently missing from current selection techniques.

$ $\\ % End line break due to AASTeX 6.3.1 errors with long acknowledgements when not using line numbers

We would like to thank the anonymous reviewer for their constructive comments which improved the final manuscript.

REH acknowledges support from the National Science Foundation Graduate Research Fellowship Program under Grant No. DGE-1746060. KNH was supported by the National Aeronautics and Space Administration (NASA) Contract NAS50210 to the University of Arizona. ADG was supported through NASA Astrophysical Data Analysis Program (ADAP) award 80NSSC23K0485.

This work makes use of color palettes created by Martin Krzywinski designed for colorblindness. The color palettes and more information can be found at \url{http://mkweb.bcgsc.ca/colorblind/}.

Observations reported here were obtained at the MMT Observatory, a joint facility of the University of Arizona and the Smithsonian Institution.
% This paper uses data products produced by the OIR Telescope Data Center, supported by the Smithsonian Astrophysical Observatory.

% This material is based upon High Performance Computing (HPC) resources supported by the University of Arizona TRIF, UITS, and Research, Innovation, and Impact (RII) and maintained by the University of Arizona Research Technologies department.

We respectfully acknowledge the University of Arizona is on the land and territories of Indigenous peoples. Today, Arizona is home to 22 federally recognized tribes, with Tucson being home to the O’odham and the Yaqui. Committed to diversity and inclusion, the University strives to build sustainable relationships with sovereign Native Nations and Indigenous communities through education offerings, partnerships, and community service.  

The Hyper Suprime-Cam (HSC) collaboration includes the astronomical communities of Japan and Taiwan, and Princeton University. The HSC instrumentation and software were developed by the National Astronomical Observatory of Japan (NAOJ), the Kavli Institute for the Physics and Mathematics of the Universe (Kavli IPMU), the University of Tokyo, the High Energy Accelerator Research Organization (KEK), the Academia Sinica Institute for Astronomy and Astrophysics in Taiwan (ASIAA), and Princeton University. Funding was contributed by the FIRST program from Japanese Cabinet Office, the Ministry of Education, Culture, Sports, Science and Technology (MEXT), the Japan Society for the Promotion of Science (JSPS), Japan Science and Technology Agency (JST), the Toray Science Foundation, NAOJ, Kavli IPMU, KEK, ASIAA, and Princeton University. 

This paper makes use of software developed for the Large Synoptic Survey Telescope. We thank the LSST Project for making their code available as free software at \url{http://dm.lsst.org}.

The Pan-STARRS1 Surveys (PS1) have been made possible through contributions of the Institute for Astronomy, the University of Hawaii, the Pan-STARRS Project Office, the Max-Planck Society and its participating institutes, the Max Planck Institute for Astronomy, Heidelberg and the Max Planck Institute for Extraterrestrial Physics, Garching, The Johns Hopkins University, Durham University, the University of Edinburgh, Queen’s University Belfast, the Harvard-Smithsonian Center for Astrophysics, the Las Cumbres Observatory Global Telescope Network Incorporated, the National Central University of Taiwan, the Space Telescope Science Institute, the National Aeronautics and Space Administration under Grant No. NNX08AR22G issued through the Planetary Science Division of the NASA Science Mission Directorate, the National Science Foundation under Grant No. AST-1238877, the University of Maryland, and Eotvos Lorand University (ELTE) and the Los Alamos National Laboratory.

Based [in part] on data collected at the Subaru Telescope and retrieved from the HSC data archive system, which is operated by Subaru Telescope and Astronomy Data Center at National Astronomical Observatory of Japan.

This publication makes use of data products from the Wide-field Infrared Survey Explorer, which is a joint project of the University of California, Los Angeles, and the Jet Propulsion Laboratory/California Institute of Technology, funded by the National Aeronautics and Space Administration.

This publication also makes use of data products from NEOWISE, which is a project of the Jet Propulsion Laboratory/California Institute of Technology, funded by the Planetary Science Division of the National Aeronautics and Space Administration.

This work has made use of data from the European Space Agency (ESA) mission
Gaia (\url{https://www.cosmos.esa.int/gaia}), processed by the Gaia
Data Processing and Analysis Consortium (DPAC,
\url{https://www.cosmos.esa.int/web/gaia/dpac/consortium}). Funding for the DPAC
has been provided by national institutions, in particular the institutions
participating in the Gaia Multilateral Agreement.

Funding for the Sloan Digital Sky Survey IV has been provided by the Alfred P. Sloan Foundation, the U.S. Department of Energy Office of Science, and the Participating Institutions.

SDSS-IV acknowledges support and resources from the Center for High Performance Computing at the University of Utah. The SDSS website is \url{www.sdss4.org}.

SDSS-IV is managed by the Astrophysical Research Consortium for the Participating Institutions of the SDSS Collaboration including the Brazilian Participation Group, the Carnegie Institution for Science, Carnegie Mellon University, Center for Astrophysics | Harvard \& Smithsonian, the Chilean Participation Group, the French Participation Group, Instituto de Astrof\'isica de Canarias, The Johns Hopkins University, Kavli Institute for the Physics and Mathematics of the Universe (IPMU) / University of Tokyo, the Korean Participation Group, Lawrence Berkeley National Laboratory, Leibniz Institut f\"ur Astrophysik Potsdam (AIP), Max-Planck-Institut f\"ur Astronomie (MPIA Heidelberg), Max-Planck-Institut f\"ur Astrophysik (MPA Garching), Max-Planck-Institut f\"ur Extraterrestrische Physik (MPE), National Astronomical Observatories of China, New Mexico State University, New York University, University of Notre Dame, Observat\'ario Nacional / MCTI, The Ohio State University, Pennsylvania State University, Shanghai Astronomical Observatory, United Kingdom Participation Group, Universidad Nacional Aut\'onoma de M\'exico, University of Arizona, University of Colorado Boulder, University of Oxford, University of Portsmouth, University of Utah, University of Virginia, University of Washington, University of Wisconsin, Vanderbilt University, and Yale University.

% GAMA is a joint European-Australasian project based around a spectroscopic campaign using the Anglo-Australian Telescope. The GAMA input catalogue is based on data taken from the Sloan Digital Sky Survey and the UKIRT Infrared Deep Sky Survey. Complementary imaging of the GAMA regions is being obtained by a number of independent survey programmes including GALEX MIS, VST KiDS, VISTA VIKING, WISE, Herschel-ATLAS, GMRT and ASKAP providing UV to radio coverage. GAMA is funded by the STFC (UK), the ARC (Australia), the AAO, and the participating institutions. The GAMA website is \url{http://www.gama-survey.org/}.

% Based on observations made with ESO Telescopes at the La Silla Paranal Observatory under programme ID 179.A-2004.

% Based on observations made with ESO Telescopes at the La Silla Paranal Observatory under programme ID 177.A-3016. 

\facilities{MMT (Hectospec), Sloan, Subaru (HSC), WISE, NEOWISE}

\software{\texttt{Astropy} \citep{collaborationAstropyCommunityPython2013}, \texttt{GELATO} \citep{GELATOv2.5.2}, \texttt{HSRED} \citep[Section 7]{fabricantHectospecMMT3002005}, \LaTeX\ \citep{lamportLaTeXDocumentPreparation1994}, \texttt{Matplotlib} \citep{hunterMatplotlib2DGraphics2007}, \texttt{NumPy} \citep{oliphantGuideNumPy2006,vanderwaltNumPyArrayStructure2011, harrisArrayProgrammingNumPy2020}, \texttt{redshifting} \citep{johnsonGalaxyQuasarFueling2018,heltonDiscoveryOriginsGiant2021,johnsonDirectlyTracingCool2022}, \texttt{SciPy} \citep{virtanenSciPyFundamentalAlgorithms2020}, \texttt{SpectRes} \citep{carnallSpectResFastSpectral2017,carnallSpectResSimpleSpectral2021}}

% \textbf{\appendix
% \restartappendixnumbering}

\bibliographystyle{aasjournal}
\bibliography{bibliography}{}

\end{document}